\documentclass[twocolumn,
               10pt,
               cleanfoot,
               ]{asme2ej}

\usepackage{graphicx}  %% for loading jpg figures
\usepackage{amsmath, amssymb, amsfonts}   % for \align 
\usepackage{enumerate}
\usepackage{chngcntr}
\usepackage{hyperref}
\usepackage{algorithm, algpseudocode}

\usepackage{amsthm}

\newcommand{\rtot}{R_{\mathrm{tot}}}
\newcommand{\qtot}{Q_{\mathrm{tot}}}
\newcommand{\itot}{I}

\theoremstyle{definition}
\newtheorem{definition}{Definition}[section]
\theoremstyle{remark}
\newtheorem{remark}{Remark}
\newtheorem{theorem}{Theorem}

\usepackage{soul,color}
\usepackage[dvipsnames]{xcolor}
\newif\ifrespondingtoreviewers
\respondingtoreviewersfalse
\ifrespondingtoreviewers
    \newcommand{\reva}[1]{\sethlcolor{cyan}\hl{#1}}
    \newcommand{\revb}[1]{\sethlcolor{yellow}\hl{#1}}
    
    \newcommand{\rev}[1]{\sethlcolor{yellow}\hl{#1}}
\else
    \newcommand{\rev}[1]{#1}
    \newcommand{\reva}[1]{#1}
    \newcommand{\revb}[1]{#1}
    
\fi

\title{Current imbalance in dissimilar parallel-connected batteries and the fate of degradation \rev{convergence}}

\author{Andrew Weng, Hamidreza Movahedi, Clement Wong, Jason B. Siegel, Anna Stefanopoulou
    \affiliation{\\
        Department or Mechanical Engineering\\
        University of Michigan, Ann Arbor, MI\\
    }
}

\begin{document}

\maketitle    

%%%%%%%%%%%%%%%%%%%%%%%%%%%%%%%%%%%%%%%%%%%%%%%%%%%%%%%%%%%%%%%%%%%%%%
\begin{abstract}
{\it 

This paper proposes an analytical framework \rev{describing how initial capacity and resistance} variability in parallel-connected battery cells may inflict additional variability or reduce variability while the cells age. We derive closed-form equations \rev{for} current and SOC imbalance dynamics within a charge or discharge cycle. These dynamics are represented by a first-order equivalent circuit model {and validated against experimental data}. To demonstrate how current and SOC imbalance leads to cell degradation, we developed a successive update scheme in which the inter-cycle imbalance dynamics update the intra-cycle degradation dynamics, and vice versa. Using this framework, \revb{we demonstrate that current imbalance can cause convergent degradation trajectories, consistent with previous reports. However, we also demonstrate that different degradation assumptions, such as those associated with SOC imbalance, may cause divergent degradation. We finally highlight the role of different cell chemistries, including different OCV function nonlinearities, on system behavior, and derive analytical bounds on the SOC imbalance using Lyapunov analysis.} \\

\noindent\normalfont Keywords: lithium-ion batteries, parallel connection, current imbalance, \rev{capacity variation, resistance variation}, degradation convergence, SEI growth, Lyapunov analysis
}
\end{abstract}

%%%%%%%%%%%%%%%%%%%%%%%%%%%%%%%%%%%%%%%%%%%%%%%%%%%%%%%%%%%%%%%%%%%%%%

\section{Introduction}

%With a steady supply of automotive-grade lithium-ion batteries, electric vehicles (EVs) captured 10\% of new passenger vehicle sales globally in 2022, a number expected to rise to 50\% by 2030 \cite{Volta_Foundation2022-wj}. Beyond EVs, batteries are also needed for energy storage systems (ESS) to enable renewable energy generation from intermittent sources such as wind and solar. ESS installations are projected to grow by 15x, exceeding 1,000 GWh by 2030 \cite{Volta_Foundation2022-wj}. Maintaining the trajectory towards full decarbonization will thus require building even more battery ``gigafactories,'' a need already recognized by industry and policy makers.
The transition to sustainable energy and transportation will require building and operating battery manufacturing factories at a gigawatt-hour scale. Among the many challenges with rapidly opening new battery factories, the question of ``how much manufacturing variability is too much?'' remains pertinent. The presence of variability in battery cell capacities and resistances is widely known \cite{Baumhofer2014-xc, Schindler2021-pn, Wildfeuer2021-xs}, the origins of which can be traced to manufacturing process variations affecting electrode-level thicknesses and loadings \cite{Kenney2012-hl, Schmidt2020-mi, Weng2023-lj}. Beyond the manufacturing of new battery packs, efforts to remanufacture second-life battery packs from aged batteries will introduce even higher variability in cell capacities and resistances \cite{Harper2019-uq, Chen2019-sb, Lai2021-xa}. Studying the effects of cell variability is thus a central question concerning both the manufacturing of new battery cells and the remanufacturing of aged battery packs.

\begin{figure*}[ht!]
\centering\includegraphics[width=\linewidth]{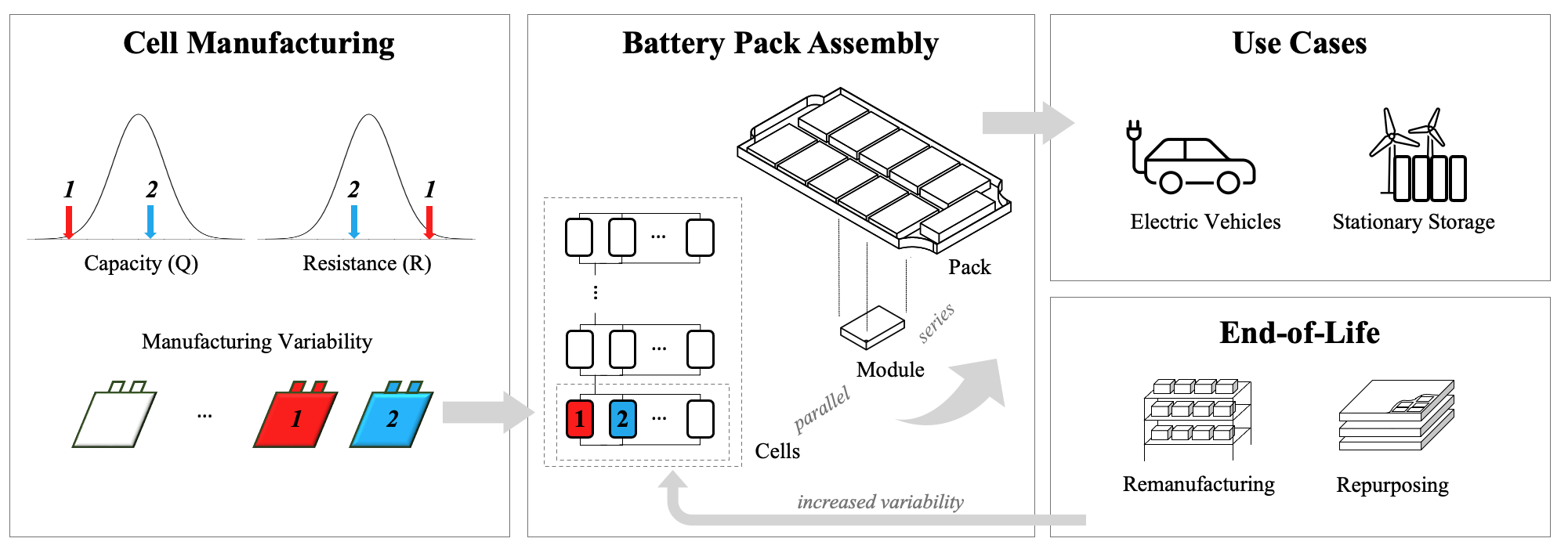}
\caption{Battery manufacturing efforts spur the need to understand the coupled dynamics of non-identical battery cells connected in parallel.}
\label{fig:concept}
\end{figure*}

When non-identical battery cells are connected in series and parallel to create a pack (see Fig. \ref{fig:concept}), the system dynamics can no longer be fully understood by studying an individual cell. In series-connected systems, \revb{for example, individual cells may be at different states of charge (SOC), but the cell having the lowest capacity is generally understood to limit the overall system capacity} \cite{Wang2021-uz, Zilberman2020-uw, Rasheed2020-ly, Feng2019-zq, Chen2023-la}. \revb{In parallel-connected systems, the currents passing through individual cells could additionally differ due to mismatches in cell internal resistances and current collection pathways. These current differences introduce another source of variability in degradation pathways, with unclear consequences over the lifetime of a battery pack. To further complicate matters, the current and SOC imbalance dynamics within parallel-connected cells are less observable in practical battery systems where individual branch currents may not be observable} \cite{Lin2020-tb}. \revb{Finally, mathematical analysis of parallel systems has also historically been hindered by the appearance of differential-algebraic equations, complicating efforts to find closed-form solutions.}

\revb{This work thus focuses on exploring the less-understood phenomena of current imbalance dynamics and variability propagation within parallel-connected battery systems.}

\subsection{Literature Review}

Existing literature on parallel-connected systems can be grouped into three approaches: experimental, simulation-based, and model-based. Experimental approaches have focused on accurately measuring the current-sharing behavior of parallel-connected battery systems using sensors including current shunts and Hall effect sensors \cite{Gong2015-mb, Brand2016-hc, Luca2021-up}. These experiments have shown that large current imbalances can persist during parallel-connected system operation \cite{Luca2021-up}. The current-sharing behavior of parallel-connected cells has since been reproduced in simulation using a variety of battery models ranging from equivalent circuit models \cite{Bruen2016-vr, Brand2016-hc, Song2021-dz, Song2022-fg} to physics-based models \cite{Reniers2023-bg}. These experimental and simulation-based approaches have enabled accurate quantification of the imbalance dynamics, especially in the context of evaluating the impact of temperature gradients \cite{Liu2019-yf, Song2021-dz, Paarmann2021-zq}. Model-based approaches have focused on developing \revb{state-space formulations enabling} state estimation \cite{Hofmann2018-gg, Zhang2020-ci, Zhang2020-dg, Drummond2021-pp}, stability analysis \cite{Li2022-kf} and steady-state analysis \cite{Brand2016-hc, Fill2021-je, Fill2018-dm}.

Despite these recent advances in measuring and modeling current imbalance in parallel-connected systems, a major gap in understanding remains: how does current imbalance affect long-term degradation behavior in parallel-connected systems? Reniers et al. \cite{Reniers2023-bg} tackled this question by simulating the degradation of an entire energy storage system, initializing each cell with different initial capacities. The authors found that initial cell-to-cell variability barely affected long-term degradation. However, the degree to which this conclusion can generalize to additional use cases and cell systems remains under-explored. \rev{Song et al.} \cite{Song2021-dz} \rev{analytically demonstrated that capacity variation can decrease over time for convex or linear degradation curves and under certain degradation model assumptions. This analysis, however, was restricted to a battery model with a linearized open-circuit voltage (OCV) function.}

\subsection{Main Contributions}

\begin{figure*}[ht!]
\centering\includegraphics[width=\linewidth]{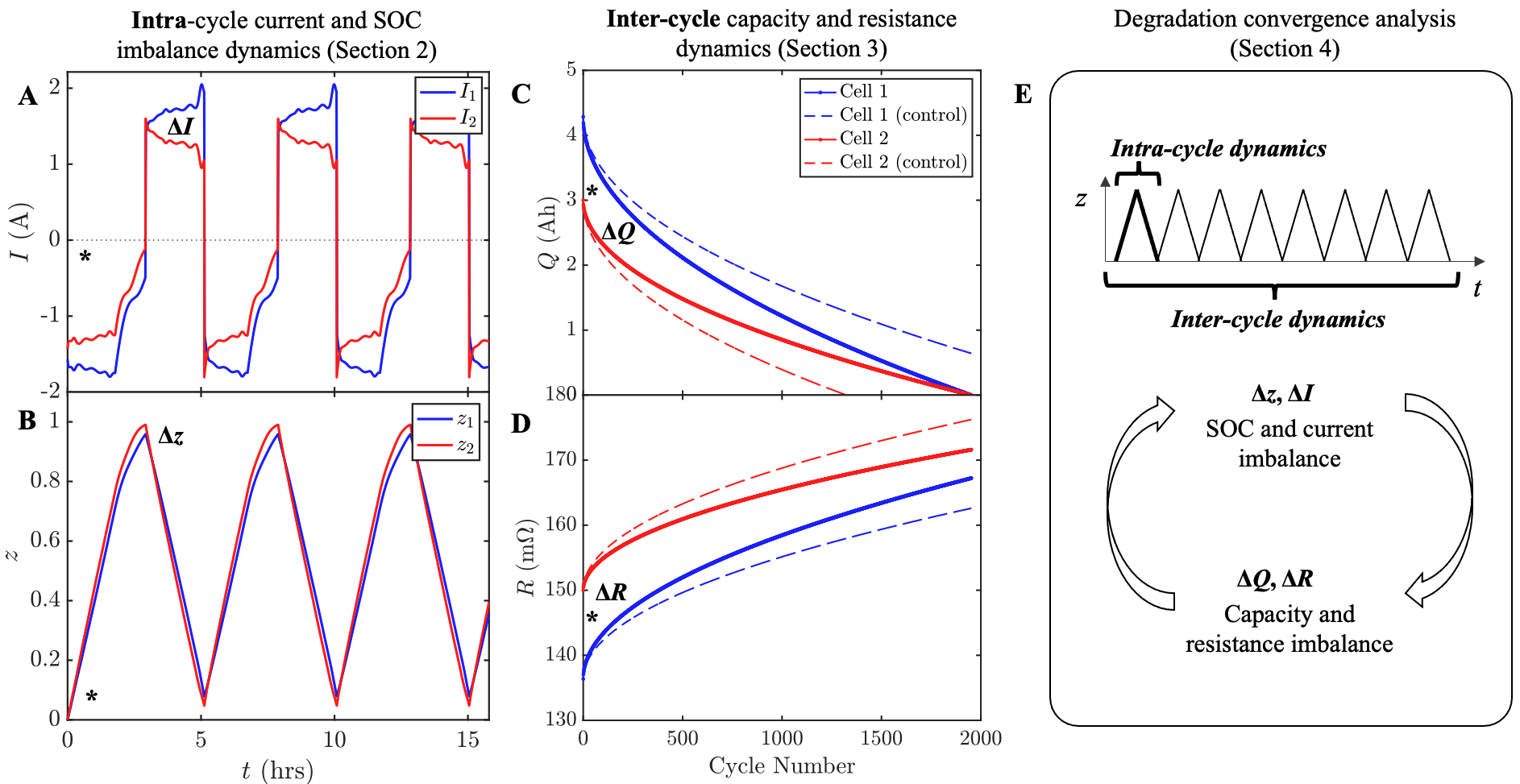}
\caption{Main contributions of this work. (A,B) Intra-cycle current and SOC imbalance dynamics are first explored in detail (Section \ref{sec:intra}). (C,D) A framework for successively updating cell capacities and resistances as a function of the intra-cycle dynamics is next proposed (Section \ref{sec:degradation}). (E) Convergence properties of degradation trajectories are explored by combining the intra-cycle dynamics with the inter-cycle dynamics from the previous sections (Section \ref{sec:intra-to-inter}). Finally, comparisons of model predictions versus experimental results (Section \ref{sec:experimental}) as well as recommended future work (Section \ref{sec:future}) are provided.}
\label{fig:inter-intra}
\end{figure*}

This work seeks to expand the analytical understanding of parallel-connected battery systems toward answering the question ``how does current and SOC imbalance within each cycle affect long-term degradation trajectories?'' (see Fig. \ref{fig:inter-intra}). Towards this goal, Section \ref{sec:intra} first develops a model of the intra-cycle \revb{(i.e. within a cycle}) battery dynamics, starting with an affine model based on equivalent circuits. We derive closed-form solutions to the system dynamics, enabling a rigorous analysis of the intra-cycle dynamics and subsequent degradation convergence behavior. We then run numerical simulations with nonlinear OCV functions for two cathode chemistries. In doing so, we assess the error introduced by the affine OCV assumption. We also analyze the stability and bounds of the SOC imbalance dynamics using Lyapunov analysis for the general case with a nonlinear OCV function. Section \ref{sec:degradation} then introduces a reduced-order degradation modeling framework enabling cycle-by-cycle updates to cell capacities and resistances as a function of SOC and current imbalance within each cycle. The work from Sections \ref{sec:intra} and \ref{sec:degradation} is combined in Section \ref{sec:intra-to-inter} to realize the simulation framework outlined in Fig. \ref{fig:inter-intra}, where the intra-cycle dynamics are used to update the inter-cycle degradation, and vice versa. Here, we highlight the importance of the underlying degradation model assumptions which ultimately determine whether degradation trajectories converge or diverge over time. Section \ref{sec:experimental} compares the modeled results to experiments, verifying the accuracy of the model-predicted intra-cycle dynamics as well as giving clues to degradation convergence/divergence for a real-world example. Section \ref{sec:future} finally suggests future research directions.

\section{Intra-Cycle Dynamics: Closed-Form Solutions}
\label{sec:intra}

We start by deriving equations describing the intra-cycle dynamics of parallel-connected systems. Section \ref{sec:affine} first develops closed-form solutions assuming an affine OCV function. Section \ref{sec:nonlinear} next discusses the system behavior with nonlinear OCV functions.

\subsection{Model Selection}

\begin{figure}[ht!]
\centering\includegraphics[width=7.5cm]{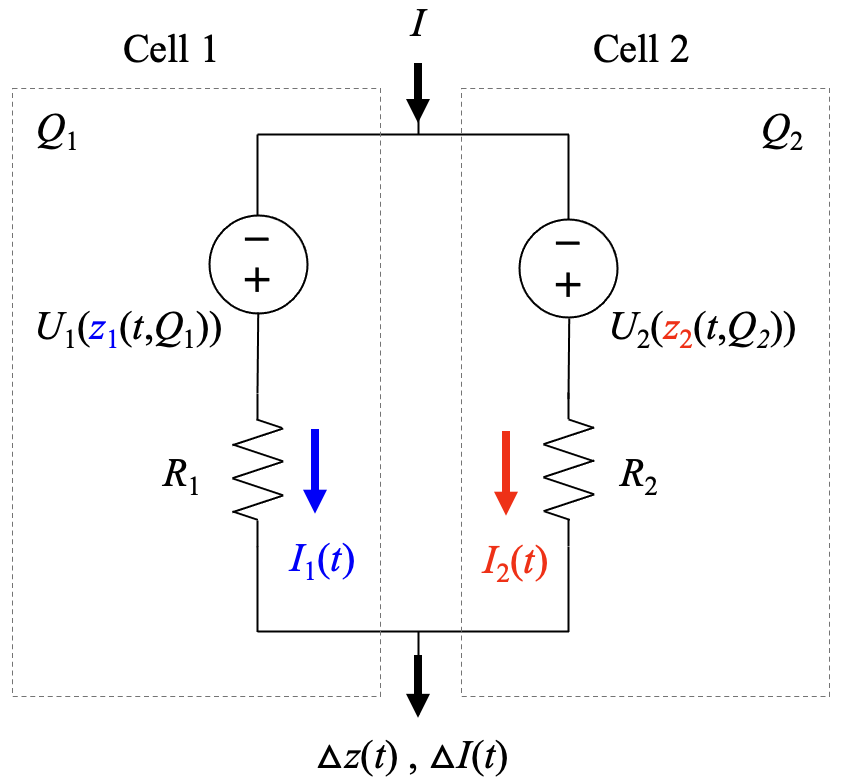}
\caption{System representation}
\label{fig:system}
\end{figure}

We chose an OCV-R model for this work (see Fig. \ref{fig:system}). This model provides the simplest representation of cell-to-cell variability in capacity and resistance. We have omitted model components that would add complexity without improving the understanding of the effect of cell-to-cell variability and degradation. Specifically, the circuit representation of the parallel-connected system \revb{omits} RC pairs and interconnect resistances. Cell resistances are also assumed to be constant parameters, \revb{though it is known that cell resistance is generally a nonlinear function} of both SOC and temperature \cite{Chen2021-ax, Weng2021-qc}. A thermal model is also omitted to first focus on studying the effect of cell capacity and resistance variability on the electrical dynamics. The analysis is finally restricted to two parallel-connected cells. The analytical framework presented here can be extended in the future to support more cells \cite{Song2021-dz}, higher degrees of nonlinearities (e.g. temperature-dependent resistances) \cite{Song2022-fg}, and more physics-based cell models such as the single-particle model (SPM) \cite{Guo2010-av, Moura2017-lg, Reniers2023-bg}.

%%%%%%%%%%%%% begin figure %%%%%%%%%%%%%%%%%
%%%%%%%%%%%%% end figure %%%%%%%%%%%%%%%%%%%

\subsection{Model Description}
\label{sec:model-description}
Fig. \ref{fig:system} shows the parallel system under consideration. $R_i$ and $Q_i$ describe the resistances and capacities of cells $i \in {1,2}$, respectively. Since modern lithium-ion have coulombic efficiencies exceeding 99\%, we assume that $R_i$ and $Q_i$ remain constant over the course of a single cycle. $U_i$ is the OCV function and is generally a nonlinear, monotonically increasing function with respect to the cell SOC $z_i$. The system input is the applied current $\itot$ which is defined to be positive on discharge. The system output is the terminal voltage, $V_t$, which is identical for all batteries due to voltage conservation, and is given by:
\begin{align}
    \label{eqn:kvl}
    V_t(t) &= U_i(z_i(t)) - I_i(t)R_i.
\end{align}
Current conservation further requires that:
\begin{equation}
    \label{eqn:kcl}
    \itot(t) = \sum_i I_i(t).
\end{equation}
Applying (\ref{eqn:kvl}) and (\ref{eqn:kcl}) for two parallel-connected cells yields the following expressions for the terminal voltage and the two branch currents:
\begin{align}
    \label{eqn:vt}
    V_t(t) &= \frac{R_1U_2(z_2(t)) + R_2U_1(z_1(t)) - R_1R_2 \itot(t)}{\rtot} \\
    \label{eqn:ib}
    I_1(t) &= \frac{-\Delta U(t) + R_2\itot(t)}{\rtot} \\
    \label{eqn:ia}
    I_2(t) &= \frac{+\Delta U(t) + R_1\itot(t)}{\rtot},
\end{align}
where $ \rtot \triangleq R_1 + R_2$ and $\Delta U(t) \triangleq U_2(z_2(t)) - U_1(z_1(t))$. The SOC dynamics for each battery are given by the integrator state:
\begin{equation}
    \label{eqn:zdot}
    \dot{z}_i(t) = -\frac{1}{Q_i} I_i(t).
\end{equation}

\begin{figure*}[ht!]
\centering\includegraphics[width=\linewidth]{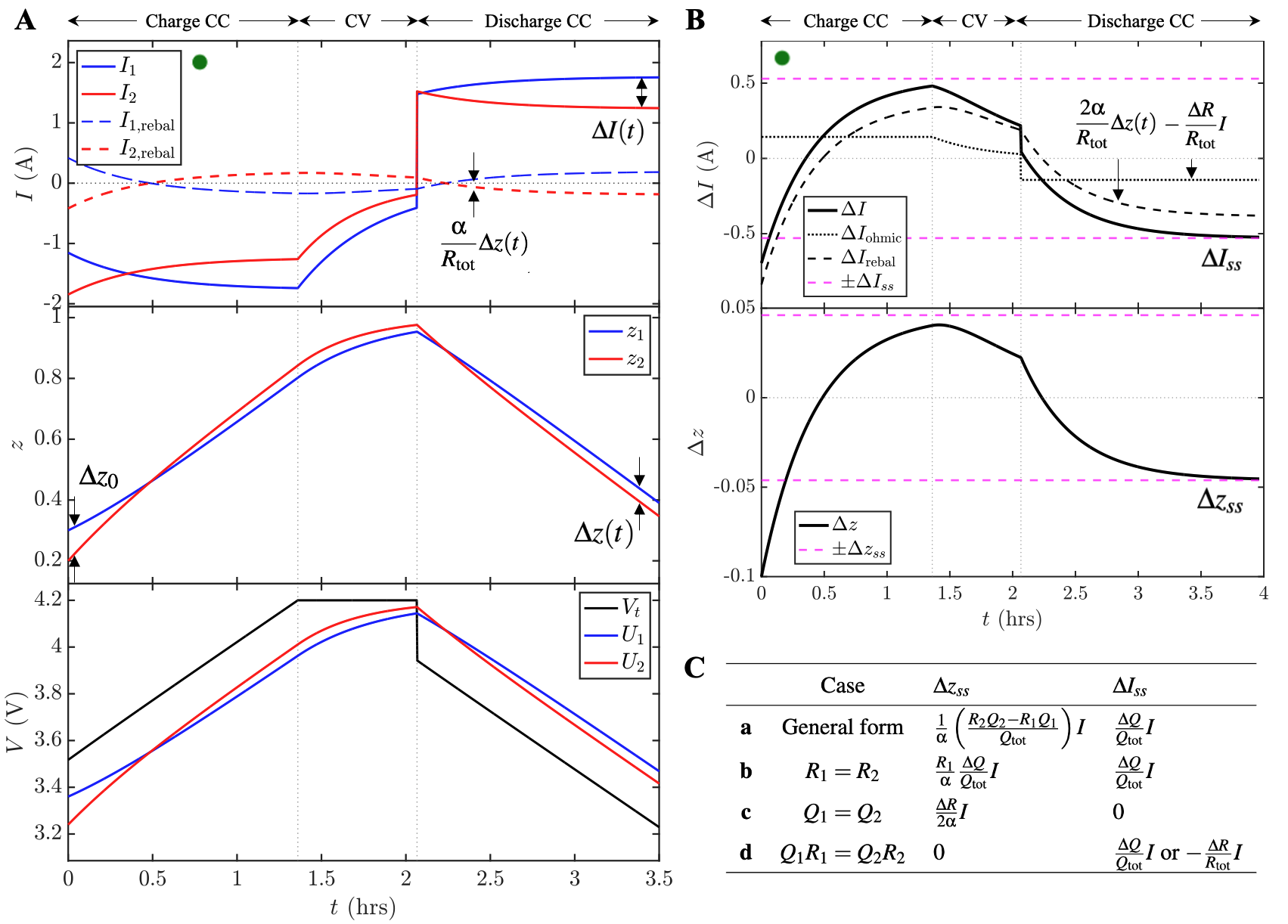}
\caption{Demonstration of the affine OCV-R system. (A) Current: $I$, SOC: $z$, and voltage: $V$, over a complete charge-discharge cycle. (B) Current and SOC imbalance. The affine OCV function parameters used are $(\alpha,\beta)=(1.2,3.0)$. The cell parameters used are $(Q_1,Q_2) = (4.3,3.0)$Ah, $(R_1,R_2)=(136,150)$m$\Omega$. This cell pairing represents an aged cell (Cell 2), with lower capacity and higher resistance, connected with a less aged cell (Cell 1). The initial condition was set to $(z_{1,0},z_{2,0}) = (0.3, 0.2)$ to highlight the effect of initial SOC imbalances on rebalancing currents. $|I|=3$A during the constant current phases, and the CV termination condition was set to $Q_2/5$A. The green markers indicate that this same set of parameters will be referenced in later figures. (C) Steady-state solutions to SOC and current imbalance.}
\label{fig:affine-1}
\end{figure*}

\subsection{The Affine OCV-R System}
\label{sec:affine}

\rev{The general behavior of the affine OCV-R system is previewed in Fig.} \ref{fig:affine-1} which serves as a reference throughout this section. This figure shows the analytical solutions throughout an entire charge and discharge cycle, including constant (CC) phases and a constant voltage (CV) phase at the top of charge. Panel A shows solutions to the branch current and cell SOC equations from Eqs. (\ref{eq:z1z2}), (\ref{eq:ibt}), (\ref{eq:iat}). Panel B shows solutions to the current and SOC imbalance equations from Eqs. (\ref{eq:dz}), (\ref{eq:di}). During the CV phase, Eqs. (\ref{eq:iacv}) and (\ref{eq:zacv}) were used. Panel C shows steady-state solutions to current and SOC imbalance.

We start by considering the case of an affine OCV function:
\begin{align} 
    \label{eqn:ocv-affine}
    U_i(z_i(t)) = \alpha z_i(t) + \beta,
\end{align}
where $\alpha, \beta > 0$. $\alpha$ is the characteristic slope of the OCV function and $\beta$ defines the minimum voltage. We note that this is the same starting point as previous works \cite{Bruen2016-vr, Fill2018-dm, Fill2021-je, Song2021-dz}. Section \ref{sec:nonlinear} later lifts this restriction and studies the model error introduced by this assumption. Here, we will also assume constant-current input unless stated otherwise.

\subsubsection{SOC Imbalance}

Combining equations (\ref{eqn:ib}-\ref{eqn:ocv-affine}) yields the following system of equations:
\begin{align}
    \label{eq:zdot-affine}
    \begin{bmatrix}
    \dot z_1 \\ \dot z_2
    \end{bmatrix}
    = \frac{\alpha}{\rtot}
    \begin{bmatrix}
    \frac{+1}{Q_1} & \frac{-1}{Q_1} \\[4pt]
    \frac{-1}{Q_2} & \frac{+1}{Q_2}
    \end{bmatrix}
    \begin{bmatrix}
    z_1 \\ z_2
    \end{bmatrix}
    - \frac{1}{\rtot}
    \begin{bmatrix}
    \frac{R_2}{Q_1} \\[4pt]
    \frac{R_1}{Q_2} 
    \end{bmatrix}
    \itot
    .
\end{align}
The solution to (\ref{eq:zdot-affine}) reads:
\begin{align}
    \label{eq:z1z2}
    \begin{bmatrix}
    z_1 \\ z_2
    \end{bmatrix}    
    = \frac{1}{\qtot}
    \begin{bmatrix}
    Q_2(1-e^{-t/\tau}) & \hphantom{x} & Q_1+Q_2e^{-t/\tau} \\
    Q_1+Q_2e^{-t/\tau} & \hphantom{x} & Q_1(1-e^{-t/\tau}) 
    \end{bmatrix}
    \begin{bmatrix}
    z_{1,0} \\ z_{2,0}
    \end{bmatrix}
    \\
    + \frac{\kappa}{\qtot}
    \begin{bmatrix}
    -Q_2(1-e^{-t/\tau})-t \\
    +Q_1(1-e^{-t/\tau})-t 
    \end{bmatrix}
    \itot,
\end{align}    
where $\qtot \triangleq Q_1 + Q_2$, and: 
\begin{align}
    \label{eq:tau}
    \tau & \triangleq \frac{\rtot}{\alpha}\left(\frac{Q_1Q_2}{\qtot}\right) & \text{(Time constant)} \\
    \label{eq:kappa}
    \kappa &\triangleq \frac{1}{\alpha}\left(\frac{R_2Q_2 - R_1Q_1}{\qtot}\right) & \text{(Input sensitivity)}.
\end{align}

\begin{remark}
The parameters $\tau$ and $\kappa$ can be physically interpreted: $\tau$ is the time constant of the system and $\kappa$ describes the input sensitivity.
\end{remark}

The SOC imbalance is defined by $\Delta z(t) \triangleq z_2(t) - z_1(t)$ and can be written by inspection from (\ref{eq:zdot-affine}) to yield:
\begin{equation}
    \label{eq:dzdot}
    \Delta \dot z(t) = -\underbrace{\frac{\alpha}{\rtot}\left(\frac{1}{Q_1} + \frac{1}{Q_2}\right)}_{\text{$1/\tau$}}\Delta z(t) +
    \underbrace{\frac{1}{\rtot}\left(\frac{R_2}{Q_1} - \frac{R_1}{Q_2}\right)}_{\text{$\kappa/\tau$}} \itot .            
\end{equation}
This is a standard linear time-invariant (LTI) system with the solution:
\begin{equation}
    \label{eq:dz}
    \Delta z(t) = \Delta z_0 e^{-t/\tau} + 
    \kappa(1-e^{-t/\tau})\itot,
\end{equation}
where $\Delta z_0 \triangleq z_{2,0} - z_{1,0}$ is the initial SOC imbalance.

\subsubsection{Branch Current Imbalance}

The branch currents for the affine OCV-R system can be written as a function of the SOC imbalance dynamics by substituting (\ref{eqn:ocv-affine}) directly into (\ref{eqn:ib}) and (\ref{eqn:ia}) to yield:

\begin{align}
    \label{eqn:I_2}
    I_1(t) &= -\frac{\alpha}{\rtot}\Delta z(t) + 
              \frac{R_2}{\rtot} \itot(t) \\
    \label{eqn:I_1}
    I_2(t) &= \underbrace{+\frac{\alpha}{\rtot}\Delta z(t)}_{I_{\mathrm{rebalance}}(t)} +                    \underbrace{\frac{R_1}{\rtot} \itot(t)}_{I_{\mathrm{ohmic}}(t)}.
\end{align}

\noindent The first terms in each equation represent the SOC rebalancing current and the second terms are due to the resistance (i.e. `Ohmic') mismatch between the two cells. Direct substitution of (\ref{eq:dz}) into (\ref{eqn:I_2}-\ref{eqn:I_1}) yields an explicit form of the branch current imbalance:

\begin{align}
    \label{eq:ibt}
    I_1(t) &= -\frac{\alpha}{\rtot}\left[\Delta z_0 e^{-t/\tau} 
              -\kappa(1-e^{-t/\tau})\itot\right]
              +\frac{R_2}{\rtot}\itot \\
    \label{eq:iat}
    I_2(t) &= \underbrace{+\frac{\alpha}{\rtot}\left[\Delta z_0 e^{-t/\tau}
              + \kappa (1 - e^{-t/\tau})\itot\right]}_{I_\mathrm{rebalance}(t)}
              + \underbrace{\frac{R_1}{\rtot}\itot}_{I_\mathrm{ohmic}(t)} .
\end{align}
The branch current imbalance $\Delta I \triangleq I_2 - I_1$ is then:
\begin{align}
    \Delta I(t) &= \frac{2}{\rtot}\Delta U(t) - \frac{\Delta R}{\rtot}I \\
                 &=\frac{2\alpha}{\rtot}\Delta z(t) - \frac{\Delta R}{\rtot} \itot \\
    \label{eq:di}
                &= \underbrace{\frac{2\alpha}{\rtot}\left[\Delta z_0e^{-t/\tau} 
    + \kappa(1 - e^{-t/\tau})\itot\right]}_{\Delta I_\mathrm{rebalance}(t)}
    - \underbrace{\frac{\Delta R}{\rtot} \itot}_{\Delta I_\mathrm{ohmic}(t)},
\end{align}
where $\Delta R \triangleq R_2 - R_1$.

A key insight from this derivation is that the input current term appears twice in Eq. (\ref{eq:di}): once as part of the SOC rebalancing term, and again as part of the Ohmic term. The applied current thus plays a dual role. First, the applied current directly contributes to the `Ohmic' current as part of the resistor network. Second, and less intuitively, the applied current also creates an internal SOC imbalance which induces its own internal SOC rebalancing current driven by the voltage difference. These two currents may flow in the same direction or in opposite directions depending on the sign of $\kappa$.

\begin{table*}[ht!]
\caption{Special cases for SOC and current imbalance in the affine OCV-R system. $\Delta Q \triangleq Q_2 - Q_1$.}
\begin{center}
\label{tbl:special-cases}
\begin{tabular}{c l c l l}
& & \\ % put some space after the caption
\hline
& Case & & $\Delta z(t)$ & $\Delta I(t) $ \\
\hline
\textbf{a} & General form & - & $\Delta z(t) = \Delta z_0 e^{-t/\tau} + \kappa(1-e^{-t/\tau})I$ & $\frac{2\alpha}{\rtot}\Delta z(t) - \frac{\Delta R}{\rtot} I$\\
\textbf{b} & Zero input & $I=0$     & $\Delta z_0 e^{-t/\tau}$ & $\frac{2\alpha}{\rtot}\Delta z_0 e^{-t/\tau}$ \\
\textbf{c} & Resistance matching & $R_1=R_2$ & $\Delta z_0 e^{-t/\tau} - \frac{R_1\Delta Q}{\alpha\qtot}(1-e^{-t/\tau})I$ & $\frac{2\alpha}{\rtot}\Delta z(t)$ \\
\textbf{d} & Capacity matching & $Q_1=Q_2$ & $\Delta z_0e^{-t/\tau} - \frac{\Delta R}{2\alpha}(1-e^{-t/\tau})I$ & $\frac{2\alpha}{\rtot}\Delta z(t) - \frac{\Delta R}{\rtot}I$ \\
\textbf{e} & `QR' matching & $Q_1R_1 = Q_2R_2 $ & $\Delta z_0 e^{-t/\tau}$ & $\frac{2\alpha}{\rtot}\Delta z_0 e^{-t/\tau} - \frac{\Delta R}{\rtot}I$ \\ 
\textbf{f} & Initial condition & $t = 0$ & $\Delta z_0$ & $ \frac{2\alpha}{\rtot}\Delta z_0$\\ 
\textbf{g} & Steady-state & $t \rightarrow \infty$ & $\frac{1}{\alpha}\left(\frac{R_2Q_2-R_1Q_1}{\qtot}\right) I \hspace{6mm} (=\kappa I$) & $\frac{\Delta Q}{\qtot}I$  \\
\textbf{h} & Maximum imbalance & $\max_t$ & $\max{(|\Delta z_0|, |\kappa I|)}$ & $\max{(|\frac{2\alpha}{\rtot}\Delta z_0|, |\frac{\Delta Q}{\qtot}I|)}$ \\ 
\hline
\end{tabular}
\end{center}
\end{table*}

\subsubsection{Potentiostatic Mode}

Since most battery charging protocols include a potentiostatic (i.e. constant voltage) hold, a complete description of a battery's charge-discharge cycle would need to consider this step. Closed-form state equations under potentiostatic mode can be derived by inverting the input and output from (\ref{eqn:vt}) to express the total current, $I_{cv}(t)$, as a function of a fixed voltage set-point $V_t = U(z=1) = \alpha + \beta$, which yields:
\begin{equation}
    \label{eq:current_during_cv}
    I_{cv}(t) = \frac{\alpha}{R_1}(z_1(t)-1) + \frac{\alpha}{R_2}(z_2(t)-1).
\end{equation}
The two terms in this equation correspond exactly to the two branch currents:
\begin{align}
    \label{eq:iacv}
    I_{i,cv}(t) &= \frac{\alpha}{R_i} (z_i(t) - 1).
\end{align}
This result can be verified by substituting (\ref{eq:current_during_cv}) into (\ref{eqn:ib}) and (\ref{eqn:ia}). The SOC dynamics can then be obtained by substituting (\ref{eq:iacv}) into (\ref{eqn:zdot}) yielding:
\begin{equation}
    \label{eq:zcv}
    \dot{z}_{i,cv} = -\frac{1}{\tau_{i,cv}}z_{i,cv} + \frac{1}{\tau_{i,cv}},
\end{equation}
where
\begin{align}
    \tau_{i,cv} \triangleq \frac{Q_iR_i}{\alpha}
\end{align}
\revb{is the characteristic time constant for SOC decay for each cell}. Eq. (\ref{eq:zcv}) is readily solved to obtain:
\begin{align}
    \label{eq:zacv}
    z_i(t) &= z_{i,cv,0}e^{-t/\tau_{i,cv}} - (e^{-t/\tau_{i,cv}} - 1).
\end{align}    

\begin{remark}
The states and time constants during the CV mode of operation are no longer coupled as was the case of CC operation: \revb{the SOC of each cell decays with its own characteristic time constant.}
\end{remark}

Finally, the SOC and current imbalance dynamics during potentiostatic mode of operation can be \revb{trivially calculated from} Eqs. (\ref{eq:iacv}) and (\ref{eq:zacv}).

\subsubsection{Affine OCV-R System Properties}
\label{sec:affine-properties}

The closed-form solutions presented here are consistent with the derivations from Refs \cite{Brand2016-hc, Fill2021-je, Li2022-kf}. \revb{However, whereas previous works mostly focused on presenting special cases (e.g. steady-state solutions), our equations are more general.} Table \ref{tbl:special-cases} summarizes how the system equations can be reduced to special cases. Useful system properties are additionally discussed below. These properties will be referenced in later sections.

\textbf{a. Stability.} The SOC imbalance system (Eq. \ref{eq:dz}), with a single negative eigenvalue with the value $\lambda=-1/\tau$, is globally exponentially stable. The current imbalance system (Eq. \ref{eq:di}) is also globally exponentially stable.

\textbf{b. Convergence Rate.} Steady-state imbalance values can be realized to within 5\% during a full charge or discharge cycle for C-rates lower than $1/3\tau$. The time constant $\tau$ depends on the cell capacities and resistances (see Eq. \ref{eq:tau}).

\textbf{c. Steady-State SOC Imbalance.} \reva{At steady-state, the SOC imbalance is:}
    \begin{align}
        \label{eq:dzss} 
        \Delta z_{ss} &= \kappa I \\
        & =\frac{1}{\alpha}\left(\frac{R_2Q_2 - R_1Q_1}{\qtot}\right) I.
    \end{align}
For the SOCs of the two cells to converge, it is thus sufficient that $Q_1R_1=Q_2R_2$. When this condition is satisfied, zero SOC imbalance is achieved at steady-state under any applied current. To understand this effect, consider the case where $Q_2 < Q_1$, $R_2 > R_1$, and $Q_1R_1=Q_2R_2$. In this case, Cell 2 sees less current due to the higher resistance. \rev{However, since Cell 2 has a lower capacity, Cell 2 experiences the same effective C-rate. Both cells thus charge at the same rate on the basis of C-Rate or SOC.}

\textbf{d. Steady-State Current Imbalance.} At steady-state, the current imbalance is
    \begin{align}
        \label{eq:diss}
        \Delta I_{ss} &= \frac{\Delta Q}{\qtot}I,
    \end{align}
where $\Delta I_{ss} \triangleq I_{ss,2} - I_{ss,1}$ and $\Delta Q \triangleq Q_2 - Q_1$. The steady-state current imbalance is thus driven by capacity mismatches, not resistance mismatches. This result can be directly recovered from Eq. (\ref{eq:di}) by taking the limit of infinite time and realizing that all of the resistance terms cancel. Note, however, that under the special case of $QR$-matching (i.e. $Q_1R_1 = Q_2R_2$), the steady-state current imbalance can be equivalently expressed as $\Delta I_{ss} = -(\Delta R/\rtot)I.$ Hence, capacity mismatch is equivalent to resistance mismatch but only under $QR$-matching.

\textbf{e. Maximum Imbalance.} If the initial SOC imbalance is zero, then the maximum SOC and current imbalances are equivalent to their steady-state values.

% \begin{table}[ht!]
% \caption{Steady-state solutions to SOC and current imbalance. $\Delta Q \triangleq Q_2 - Q_1$.}
% \begin{center}
% \label{tbl:ss}
% \begin{tabular}{c c c c}
% \hline
%  & Case & $\Delta z_{ss}$ & $\Delta I_{ss}$  \\
% \hline
% \textbf{a} & General form & $\frac{1}{\alpha}\left(\frac{R_2Q_2 - R_1Q_1}{\qtot}\right)I$ & $\frac{\Delta Q}{\qtot} I$ \\ 
% \textbf{b} & $R_1 = R_2$ & $\frac{R_1}{\alpha}\frac{\Delta Q}{\qtot}I$ & $\frac{\Delta Q}{\qtot}I$ \\
% \textbf{c} & $Q_1=Q_2$ & $\frac{\Delta R}{2\alpha}I$ & $0$ \\
% \textbf{d} & $Q_1R_1 = Q_2R_2$ & 0 & $\frac{\Delta Q}{\qtot}I$ or $-\frac{\Delta R}{\rtot}I$ \\
% \end{tabular}
% \end{center}
% \end{table}

\subsection{The Nonlinear OCV-R System}
\label{sec:nonlinear}

The affine OCV-R model derived in Section \ref{sec:affine} provided an analytically tractable description of imbalance dynamics. But how accurate is this model? This section explores this question by revisiting the OCV-R model, this time relaxing the affine OCV assumption. Section \ref{sec:nonlinear-simulations} takes a numerical approach to explore the model behavior under nonlinear OCV functions for two popular battery chemistries. Section \ref{sec:nonlinear-bounds} then takes an analytical approach to bound the maximum SOC imbalance under arbitrary, nonlinear current inputs.

\subsubsection{Numerical Simulations}
\label{sec:nonlinear-simulations}

\begin{figure}[ht!]
\centering\includegraphics[width=\linewidth]{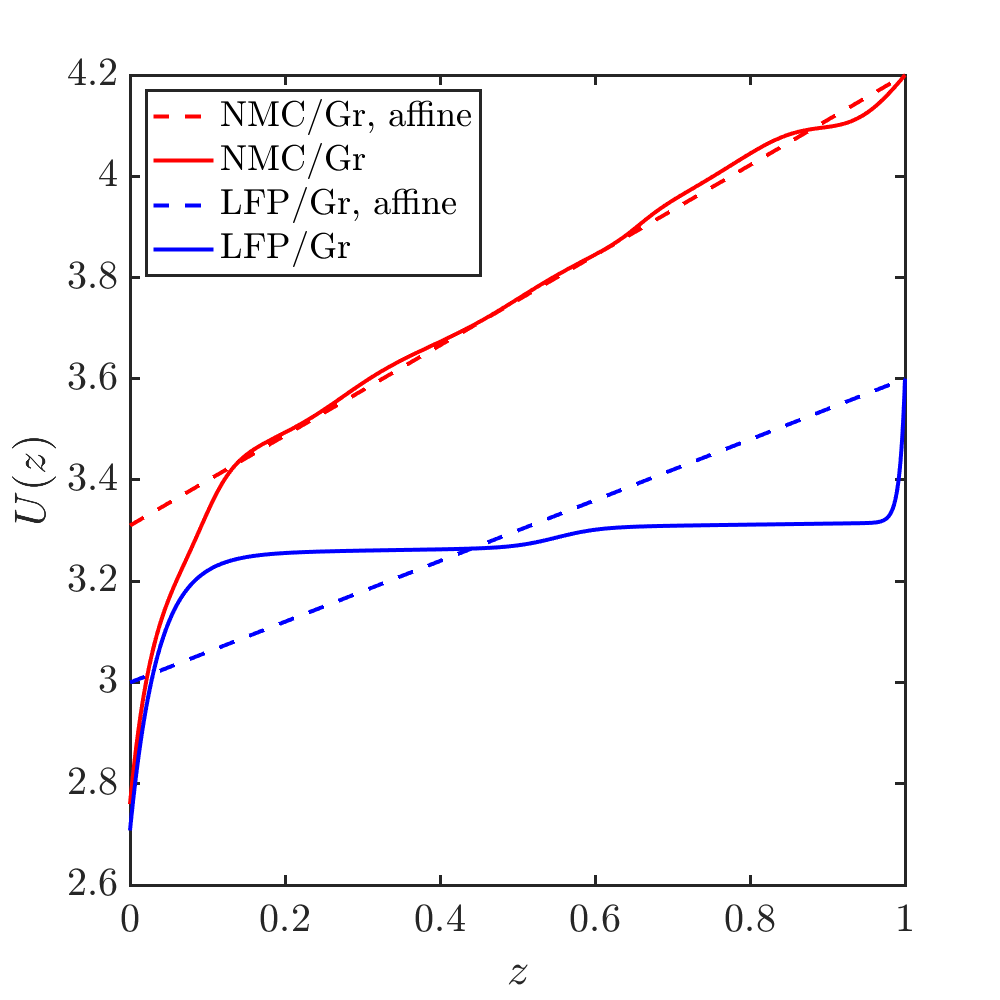}
\caption{Nonlinear OCV functions used for this work. The LFP/Gr curve was adapted from Prada et al \cite{Prada2013-zz}. The NMC/Gr curve was adapted from Chen et al \cite{Chen2020-tg}. The affine approximation to the LFP/Gr curve used $(\alpha,\beta)=(0.6, 3.0)$. The affine approximation to the NMC/Gr curve used $(\alpha,\beta)=(0.89,3.31)$.}
\label{fig:nonlinear-ocv}
\end{figure}

Here, we revert to the general case of the OCV function $U(z)$ being a nonlinear, monotonically increasing function (Section \ref{sec:model-description}). Fig. \ref{fig:nonlinear-ocv} summarizes the OCV functions used in this section. We focus on studying two OCV functions that are characteristic of two common battery cathode materials: nickel manganese cobalt cathode (NMC) \cite{Chen2020-tg} and lithium iron phosphate (LFP) \cite{Prada2013-zz}. Both cathodes were paired with conventional graphite (Gr) anodes. Affine OCV functions were also defined for comparison purposes. The nonlinear system was numerically solved by discretizing the state equations from Section \ref{sec:model-description} using a forward difference scheme with a 1-second timestep. CV mode of operation was simulated by inverting the input and output from Eq. (\ref{eqn:vt}). The code used to generate the simulation is available at \href{https://github.com/wengandrew/current-imbalance}{https://github.com/wengandrew/current-imbalance}.

Fig. \ref{fig:nonlinear-combined} compares example simulation results using the NMC/Gr and LFP/Gr OCV curves. Each simulation consisted of five back-to-back charge-discharge cycles. A CV phase with a termination condition of $Q_2/5$ was included at the end of each charge, but not at the end of each discharge. Analytical solutions for the affine OCV case were included for the first cycle only for comparison purposes. Since the duration of the CC charge phase was generally non-identical between the affine and nonlinear cases, the time vectors were realigned using the start and end of the CV phase as reference points.

\subsubsection{Nonlinear System Behavior}
 
 The nonlinear OCV functions drive nonlinear oscillations in current and SOC imbalance. \revb{Meanwhile, t}he affine OCV system only predicts exponentially-decaying behavior (Section \ref{sec:affine}). The higher SOC imbalance in the LFP/Gr system can be attributed to the flatness of the LFP/Gr OCV function which suppresses the SOC re-balancing current (Eq. \ref{eq:di}). SOC imbalance therefore accumulates in LFP/Gr systems until one of the cells reaches an inflection point in the OCV function. \revb{The majority of SOC re-balancing in LFP/Gr systems thus occurs as one of the cells approaches 100\% SOC.}

\subsubsection{Nonlinear vs Affine Model Accuracy}

\textit{NMC/Gr.} The nonlinear NMC/Gr solution deviates from the affine solution the most at low and high SOCs, corresponding to regions where the slope of the OCV function $dU/dz$ deviates the most from the affine approximation $\alpha$. Overall, the affine solution provides only an approximation to the nonlinear behavior and fails to capture the localized perturbations to current imbalance due to $dU/dz$. The appropriateness of using affine dynamics to model the nonlinear system will depend on the accuracy requirements of the specific use case.

\begin{figure*}[ht!]
\centering\includegraphics[width=\linewidth]{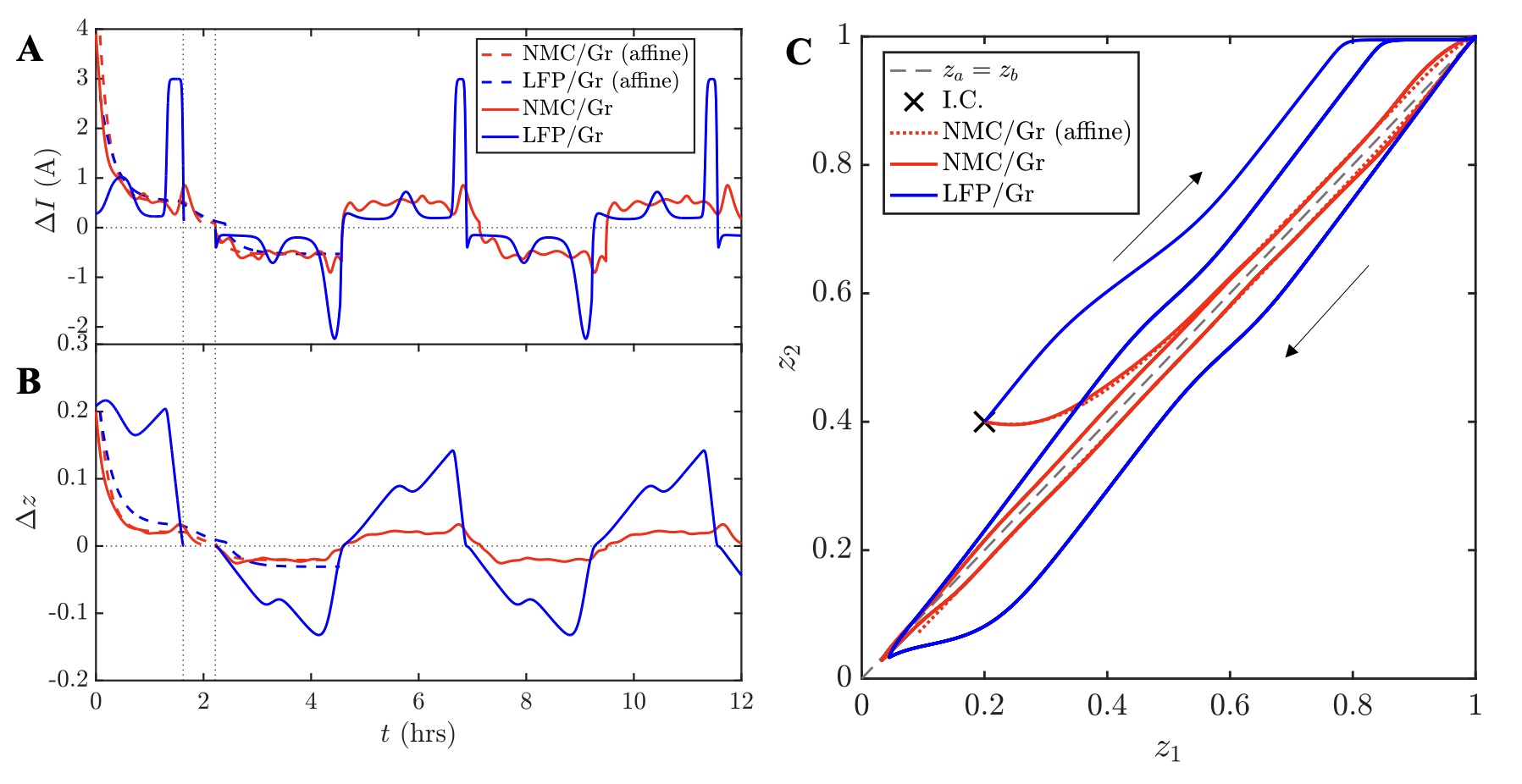}
\caption{Intra-cycle current (A) and SOC (B) imbalance dynamics of the OCV-R system with nonlinear OCV functions. Vertical dotted lines denote the start and end of the charge CV phase. (C) Corresponding phase orbits in the $(z_1,z_2)$ plane over a period of five complete charge-discharge cycles. Parameters used were $(Q_1, Q_2)=(4.28,3.00)$Ah, $(R_1, R_2)=(45.5, 50.0)$m$\Omega$, $(z_{1,0},z_{2,0}) = (0.2, 0.4)$, $|I|=3$A during the constant current phases, and with a CV termination of $Q_2/5$ A.}
\label{fig:nonlinear-combined}
\end{figure*}

\textit{LFP/Gr.} The current and SOC imbalance observed in the nonlinear LFP/Gr system is greater than those in the NMC/Gr system. Affine approximations to the LFP/Gr curves fail to capture the dynamics of the LFP/Gr system whose behavior is dominated by transitions between low and high-sloping regions in the OCV function. \rev{For this system, it may be more appropriate to use piecewise-linear OCV functions to approximate the nonlinear OCV function, which we leave for future work.}

\begin{remark} The results highlighted in Fig. \ref{fig:nonlinear-combined} depend on the input assumptions, particularly the values of $(Q_2,R_2)$ and the input magnitude $I$. The accuracy of the affine model and the behavior of the nonlinear OCV systems should thus be considered on a case-by-case basis.
\end{remark}

\subsubsection{Nonlinear System Stability}

Fig. \ref{fig:nonlinear-combined}C shows a phase portrait of the SOC dynamics over five consecutive charge-discharge cycles. In both the affine and the nonlinear cases, the system converges to a stable orbit after just one complete charge-discharge cycle. This result holds for both NMC/Gr and LFP/Gr curves, although the hysteresis gap between charge and discharge is higher for LFP/Gr. \revb{This result highlights the stability of the system under nonlinear OCV functions and that convergence to a stable orbit can be realized within a few cycles.}

\subsubsection{$QR$-Matching Nullifies Nonlinearities}
\label{sec:qr-matching}

\begin{figure*}[ht!]
\centering\includegraphics[width=\linewidth]{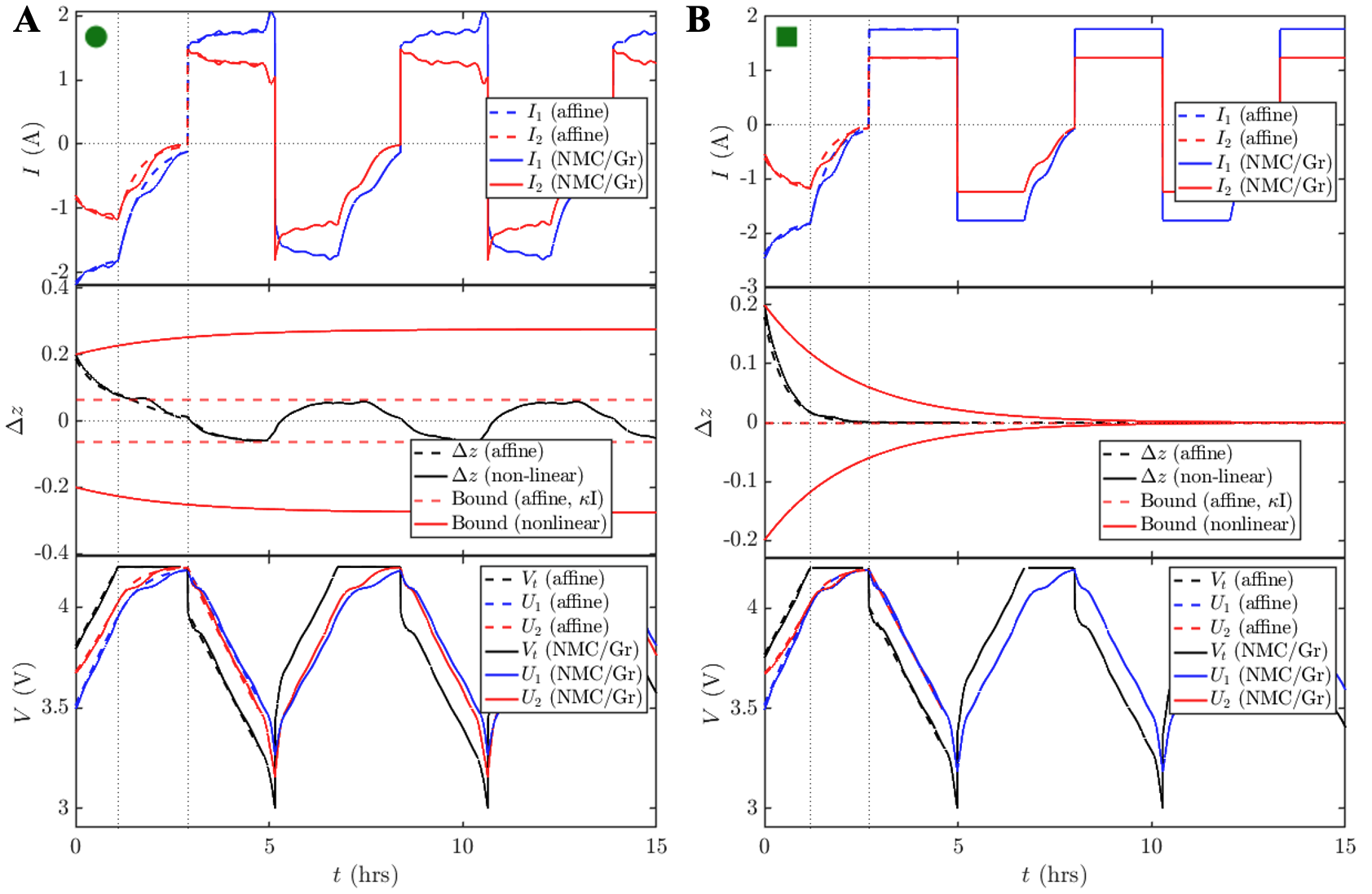}
\caption{\reva{OCV-R model system dynamics with nonlinear OCV functions} representing an NMC/Gr system. In both (A) and (B), $(Q_2,R_2)=(3$Ah$,150$m$\Omega$), $(z_{1,0},z_{2,0}) = (0.4, 0.2)$, $|I|=3$A during the constant current phases, and with a CV termination of $Q_2/5$ A. (A) uses $(Q_2/Q_1,R_2/R_1)=(0.7, 1.1)$. This pairing represents a typical scenario in which an aged cell (Cell 2), with lower capacity and higher resistance, is paired with a less aged cell (Cell 1). (B) uses $(Q_2/Q_1,R_2/R_1)=(0.7, 1.43)$. This pairing also represents a typical scenario with an aged cell paired with a less aged cell, but this pairing additionally satisfies the `$QR$-matching' condition, i.e. $Q_1R_1 = Q_2R_2$. Under this condition, the SOC imbalance dynamics become insensitive to the input current and exponentially decay to zero. The corresponding imbalance dynamics also become driven purely by resistance differences in the absence of SOC re-balancing currents. After the initial SOC imbalance fades, this system begins to behave identically to the affine OCV system, despite the presence of the nonlinear OCV function. See Section \ref{sec:qr-matching} for a complete discussion.} 
\label{fig:nonlinear-nmc}
\end{figure*}

Here, we highlight a peculiar property of the nonlinear system: when $Q_1R_1 = Q_2R_2$, the nonlinear oscillations in the SOC and currents are nullified. We will refer to this condition as `$QR$-matching'. Fig. \ref{fig:nonlinear-nmc} \revb{shows this effect by comparing} the nonlinear dynamics for the NMC/Gr system under two specific cases. Case A (circle) simulates two cells with capacities and resistance values identical to those from Fig. \ref{fig:affine-1}. \revb{Case B (square) chooses resistance and capacities satisfying $Q_1R_1 = Q_2R_2$. Case A does not satisfy the $QR$-matching condition, while Case B does.} (See Fig. \ref{fig:nonlinear-lfp} for the same plot but with LFP/Gr chemistry. The following discussion holds for both NMC/Gr and LFP/Gr chemistries.)

Panel B shows that\revb{, under $QR$-matching}, the current oscillations disappear after one cycle, \revb{and the affine solution begins to overlap the nonlinear solution. Thus, after one cycle, considerations for the nonlinear OCV dynamics appear to be no longer needed. To explain this phenomenon, consider the fact that, under $QR$-matching,} $\kappa=0$ in Eq. \ref{eq:dz}, so the SOC imbalance decays to zero under any input current. With zero SOC imbalance, the SOC re-balancing current in Eq. (\ref{eq:di}) also disappears. Current imbalance thus becomes purely driven by the resistance difference between the two cells, which does not depend on the SOC or OCV dynamics.

Overall, under $QR$-matching, the system appears to be analyzable without considering the nonlinear effects introduced by the OCV function. This result has practical implications since $QR$-balancing can describe certain scenarios in which aged cells are mixed with fresh cells. $QR$-balancing can also describe a scenario in which two cells having different electrode areas that are otherwise identical are connected in parallel.

\subsubsection{Stability and Analytic Bounds for SOC Imbalance}
\label{sec:nonlinear-bounds}

The previous sections showed that\revb{, with the exception of the $QR$-matching condition, the affine OCV-R model may fail to capture transient behavior in SOC and current imbalance, especially for systems having OCV functions with widely-varying slopes such as LFP/Gr. OCV function nonlinearities should thus be generally taken into account for accurate estimates of SOC and current imbalance. However, a nonlinear OCV function makes it difficult to find closed-form solutions. Yet, despite the absence of closed-form solutions, we may still attempt to use analytical methods to derive bounds on the maximum imbalance.}

To derive imbalance bounds, we leverage concepts of input-output and $\mathcal{L}$-stability from Khalil \cite{Khalil2002}. Note that the imbalance system is not strictly asymptotically stable, but only partially asymptotically stable, since the SOCs can asymptotically approach any value between 0 and 1. We thus also leverage definitions from Haddad et al. \cite{Haddad2008} which apply to partially asymptotically stable systems. The main result is summarized below.

\begin{theorem}
\label{eq:theorem}
\reva{If the following condition is satisfied:}
\begin{equation}
    \label{eq:condition}
    \mathrm{max}(|I(t)|) \leq |\frac{\mathcal{A}k_1}{\mathcal{B}}|,
\end{equation}
then $\Delta z(t)$ satisfies the following bounds:
\begin{align}
    \label{eq:linfbound}
    \max(||\Delta z||) &\leq |\Delta z(0)| e^{k_1\mathcal{A}t} + |\frac{\mathcal{B}}{\mathcal{A}k_1}|\max(|I(t)|)(1-e^{k_1\mathcal{A}t})
\end{align}
where:
\begin{align}
    \label{eq:k1k1}
    k_1 &= \mathrm{min}\left(\frac{\partial U(z)}{\partial z}\right) \\
    \mathcal{A} &= -\frac{1}{\rtot}\left(\frac{1}{Q_1} + \frac{1}{Q_2}\right) \\
    \mathcal{B} &= \frac{1}{\rtot}\left(\frac{R_1}{Q_2} - \frac{R_2}{Q_1}\right) 
\end{align}
\end{theorem}

\begin{remark}
Eq. (\ref{eq:linfbound}) provides an ${l}_2$-vector norm bound. A signal-norm bound can similarly be calculated.
\end{remark}

A complete derivation of this result is provided in Appendix A. 

\subsubsection{Application of SOC Imbalance Bounds}

To understand the utility of the nonlinear imbalance bounds, we applied the bounds to the cases shown in Fig. \ref{fig:nonlinear-nmc}. Horizontal lines drawn in the second rows show infinity-norm bounds (\ref{eq:linfbound}) on the SOC imbalance. These bounds were computed for the cases shown in Panels A and B which both satisfy condition (\ref{eq:condition}). These bounds were compared against the affine solution to the steady-state SOC imbalance ($\pm\kappa I$), which \revb{we interpret as} `affine bounds.' Panel A shows that these affine bounds were exceeded initially since \revb{they fail to capture the} effect of initial SOC imbalance. However, after the initial transient response decays, the affine bounds successfully bounded the current imbalance for the remainder of the simulation. By comparison, the nonlinear bounds gave correct, albeit conservative, bounds on the maximum SOC imbalance. In Panel B, the affine bound again could not capture the initial SOC imbalance but trivially predicted the SOC imbalance at steady-state which decays to zero. Meanwhile, the nonlinear bound was able to bound both the initial SOC imbalance and capture the decay of the SOC imbalance towards zero at steady state.

This demonstration shows that the nonlinear bounds (\ref{eq:linfbound}) \revb{correctly but weakly bound the SOC imbalance. The weakness of the bounds can be attributed to the fact that $k_1$, the minimum slope of the OCV function} (\ref{eq:k1k1}), must be large in order for the bound to be tight. Yet, for most practical lithium-ion battery chemistries, OCV functions often have regions with shallow slopes, so $k_1$ is generally small. Thus, battery chemistries having very flat OCV curves such as LFP may not be able to derive utility from these bounds (see Fig. \ref{fig:nonlinear-lfp}). By comparison, the affine bound was surprisingly effective at providing bounds on the SOC imbalance after the initial transient response decays, at least for the NMC/Gr system. However, for the LFP/Gr system, the affine bounds also failed to bound the SOC imbalance since the rapid change in the OCV slope at the top of charge takes this system out of steady-state (see Fig. \ref{fig:nonlinear-lfp}).

\section{Inter-Cycle Degradation: Successive Update Scheme}
\label{sec:degradation}

Updating cell capacities and resistances based on the intra-cycle current and SOC dynamics derived in Section \ref{sec:intra} requires a degradation model. This model specifically needs to allow for updates to the degradation states (i.e. capacity and resistance) as a function of time-varying parameters such as currents and SOCs which could change cycle-by-cycle. Empirical models, which rely on experimental curve fits to degradation data, cannot be used for this work since these models assume that the intra-cycle dynamics (e.g. C-rates, depths of discharges) remain fixed \cite{Smith2021-qv}. More physics-based approaches are thus needed. Such approaches may provide electrode-level state variables, such as solid-phase lithium concentrations and reaction current densities, which can be more directly tied to relevant degradation modes such as loss of lithium inventory and loss of active material \cite{OKane2022-fj, Birkl2017-yq}. Reniers et al. \cite{Reniers2023-bg}, for example, used a single particle model (SPM) to represent the intra-cycle battery dynamics for each cell in a system of series and parallel-connected cells, and successfully coupled these dynamics to an `inter-cycle' degradation model based on SEI growth and electrode particle cracking.

Here, we develop a simplified, semi-empirical representation of battery degradation enabling cycle-by-cycle cell capacity and resistance updates as a function of current and SOC imbalance. The model is semi-empirical in that degradation variables are restricted to full cell-level quantities (i.e. full cell capacities and resistances). Since this work focuses on making an elementary connection between the intra-cycle dynamics of parallel-connected systems to battery degradation, we have chosen the simplest representation of degradation. Provisions for electrode-level state variables (e.g. solid-phase lithium concentrations and interfacial potentials) have thus been omitted. The methodologies presented here, however, can be applied to more realistic degradation models and higher-fidelity models of battery dynamics, which is left for future work.

\subsection{Incremental Capacity Loss Model Formulation}

\label{sec:incremental}

We start by considering discrete updates to cell capacity by writing down the incremental capacity loss at each cycle. The capacity of cell $i$ at any given cycle number is:
\begin{align}
    \label{eq:qupdate}
    Q_{i,n} &= Q_{i,0} - L_{i,n},
\end{align}
where $n \in {1, 2, ...}$ is the cycle number, $Q_{i,0}$ is the initial cell capacity, and $L_{i,n}$ is the total capacity lost at the end of the $n$th cycle. $L_{i,n}$ is determined by multiple physical degradation phenomena including loss of lithium inventory and loss of active material \cite{Birkl2017-yq}. Here, we adopt a generic form of the loss equation presented in Smith et al. \cite{Smith2021-qv} for solid-electrolyte interphase (SEI) growth \cite{Pinson2012-wt}:
\begin{equation}
    \label{eq:dxdt_sei}
    \frac{dL_i(t)}{dt} = r_i(t) \cdot p \cdot \left(\frac{r_i(t)}{L_i(t)}\right)^{\frac{1-p}{p}},
\end{equation}
where $L_i$ is the total capacity lost at time $t$, $r_i(t)$ is a time-dependent reaction rate constant and $p$ is an exponential factor. $p$ is allowed to vary between 0.5 in the case of pure diffusion-limited SEI growth, and 1.0 in the case of pure reaction-limited SEI growth. By taking a constant reaction rate, Eq. (\ref{eq:dxdt_sei}) evaluates to the familiar form of $L_i(t) = r_it^p$. However, this equation cannot be directly used if $r$ changes cycle-by-cycle, which would lead to discontinuities in the degradation curve. 

\begin{figure}[ht!]
\centering\includegraphics[width=7cm]{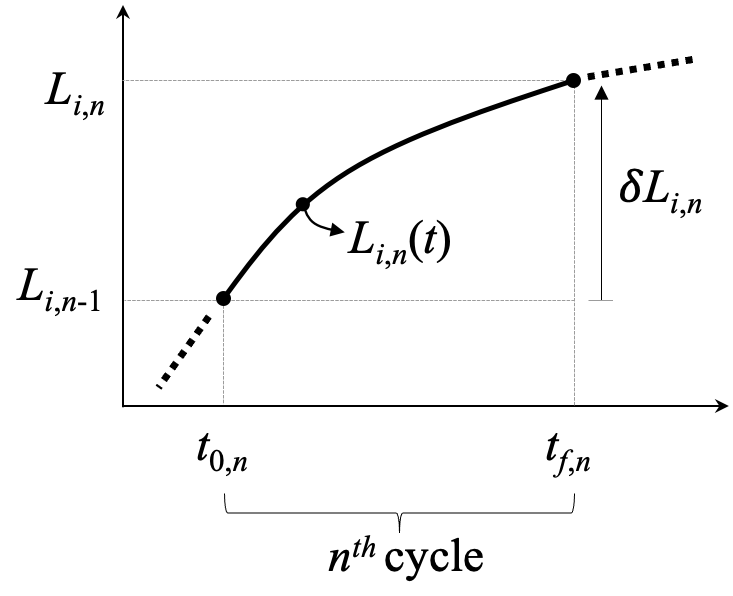}
\caption{Methodology for calculating the incremental capacity loss at cycle $n$. $L_{i,n}$ represents the capacity lost at cycle $n$ and is a discrete quantity. $L_{i,n}(t)$ is a continuous variable describing the instantaneous capacity loss value during cycle $n$ and at time $t$, and is valid over the domain $t_{0,n}<t<t_{f,n}$.}
\label{fig:degradation-calculation}
\end{figure}

We therefore develop an integral form of Eq. (\ref{eq:dxdt_sei}) to describe the incremental capacity lost over a single cycle $n$, using Fig. \ref{fig:degradation-calculation} as a guide. In this formulation, $r_i$ is allowed to vary cycle-to-cycle and within each cycle. \reva{The integration is performed through separation of variables, taking the initial condition to be $(t_0,L_{i,n-1}$), where $t_0$ is the time at the start of the $n$th cycle and $L_{i,n-1}$ is the total capacity lost immediately before the start of the $n$th cycle. Performing this integration yields:}
\begin{equation}
    L_{i,n}(t) = \left(\int_{t_{0,n}}^{t}r_{i,n}(\tau)^{1/p}d\tau + L_{i,n-1}^{1/p}\right)^p,
\end{equation} where $r_{i,n}$ now denotes the reaction rate, valid over cycle $n$. $L_{i,n}(t)$ is valid over the domain $t_{0,n}<t<t_{f,n}$, where $t_{0,n}$ and $t_{f,n}$ denote the start and the end time of the $n$th cycle. The total capacity lost at the end of the $n$th cycle is then $L_{i,n} \triangleq L_{i,n}(t_{f,n})$. The incremental capacity loss during cycle $n$ is:
\begin{align} 
    \delta L_{i,n} &\triangleq L_{i,n} - L_{i,n-1} \\
    \label{eq:delta_l}
      &= \left(\int_{t_{0,n}}^{t_{f,n}}r_{i,n}(\tau)^{1/p}d\tau + L_{i,n-1}^{1/p}\right)^p - L_{i,n-1}.
\end{align}
If $r_{i,n}$ is further assumed to be constant over the cycle, then the expression simplifies to:
\begin{equation}
    \label{eq:delta_l_simplified}
    \delta L_{i,n} = \left(r_{i,n}^{1/p}\Delta t_n + L_{i,n-1}^{1/p}\right)^p - L_{i,n-1},
\end{equation}
where $\Delta t_n \triangleq t_{f,n} - t_{0,n}$.

Eqs. (\ref{eq:delta_l}) and (\ref{eq:delta_l_simplified}) clarify the fact that the capacity lost at each cycle depends on how much capacity was lost previously. This result can be physically interpreted for $p<1$ which corresponds to a self-limiting SEI growth mechanism in which more capacity lost implies slower reaction rates, and hence lower capacity loss rates \cite{Pinson2012-wt}. A similar interpretation of capacity loss is discussed in the context of path-dependent aging in Karger et al. \cite{Karger2022-kt}. 

To model resistance growth, we adopt a simple approach by considering the cell resistance at each cycle to be:
\begin{equation}
    \label{eq:rupdate}
    R_{i,n} = R_{i,0} + G_{i,n},
\end{equation}
where $R_{i,0}$ is the initial cell resistance and $G_{i,n}$ is the total resistance growth at the end of the $n$th cycle. We assume that $G_{i,n}$ is related to $L_{i,n}$ according to
\begin{equation}
    \label{eq:resistance}
    G_{i,n} = \lambda_1 L_{i,n} + \lambda_2,
\end{equation}
where $\lambda_1>0$ is a proportionality factor that describes the lithium-consuming SEI film growth process which leads to resistance growth \cite{Yang2017-uf} and $\lambda_2>0$ describes resistance growth contributions occurring independently from lithium-consuming processes such as SEI growth (i.e. film growth in layered oxide cathodes \cite{Abraham2005-fd}).

\begin{figure}[ht!]
\centering\includegraphics[width=\linewidth]{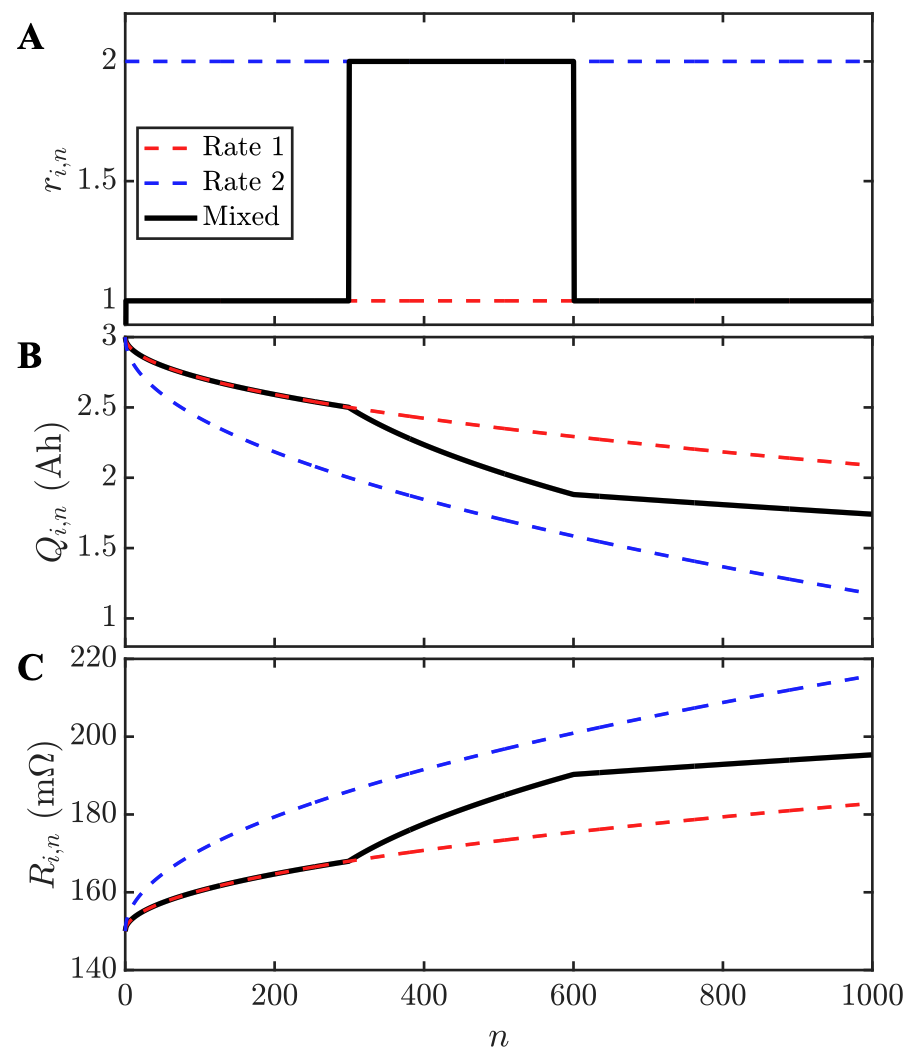}
\caption{Demonstration of the incremental capacity loss update model which supports cycle-by-cycle updates to the reaction rate $r$. \revb{(A) Cycle-dependent reaction rate assumptions for each degradation trajectory. (B) Capacity fade over cycles. (C) Resistance growth over cycles.}}
\label{fig:degradation-sei}
\end{figure}

Fig. \ref{fig:degradation-sei} demonstrates how this model formulation can be used to simulate dynamic capacity fade trajectories on a single cell. In this simulation, $\delta L_i$ is calculated at each cycle for an arbitrary cell $i$ via successive updates according to Eq. (\ref{eq:delta_l_simplified}). The cell capacity is then updated according to Eq. (\ref{eq:qupdate}). For demonstration purposes, $r_{i}$ is set to 2 between 300 and 600 cycles and is otherwise set to 1 (Panel A). This trajectory is compared to two control trajectories in which $r_{i}$ remain invariant throughout the simulation. \revb{The control trajectories bounded the dynamic capacity fade trajectory but neither correctly predicted the final cell capacity (Panel B). A similar observation follows for the resistance growth trend (Panel C), which was updated according to Eq.} (\ref{eq:rupdate}).

\subsection{Coupling Reaction Rates to Imbalance Dynamics}

The incremental capacity loss model derived here can be used to study the effect of changing use conditions, irrespective of whether the change is due to external factors (e.g. customer use patterns) or internal factors (e.g. parallel-connected battery dynamics). In both cases, the effect of changing use conditions can be represented by treating the current and SOC dynamics as modifications to $r_i(t)$:
\begin{equation}
    \label{eq:rin-general}
    r_{i,n}(t) = g(z_{i,n}(t), I_{i,n}(t)),
\end{equation}
where $z_{i,n}(t)$ and $I_{i,n}(t)$ are the SOC and current for cell $i$ and cycle $n$. Since the imbalance values will generally change over life, $r_{i,n}(t)$ will take on different values over each cycle. In this manner, the incremental degradation in each cell at each cycle, $\delta L_{i,n}$, becomes coupled with the intra-cycle dynamics of that cell. 

To resolve $r_{i,n}(t)$, expressions for $z_{i,n}(t)$ and $I_{i,n}(t)$ are needed. Fortunately, we have already developed analytical expressions for $z_{i,n}(t)$ and $I_{i,n}(t)$ in Section \ref{sec:intra}. We next explore model simplifications to $g(z_{i,n}(t), I_{i,n}(t))$ and what these simplifications imply about degradation convergence.

\section{Degradation Convergence Analysis}
\label{sec:intra-to-inter}

This section highlights how the intra-cycle dynamics developed in Section \ref{sec:intra} can be combined with the inter-cycle capacity loss formulation developed in Section \ref{sec:degradation} to analyze and simulate whether capacity degradation converges or diverges over the course of many cycles. 

We will focus specifically on the case of a fresh cell connected in parallel with an aged cell. Taking Cell 1 to be the fresh cell and Cell 2 to be the aged cell, we thus impose the constraints that $Q_2 < Q_1$ (the aged cell has lost capacity) and $R_2 > R_1$ (the aged cell has increased resistance). Fig. \ref{fig:affine-1} showed an example of one such system.

\begin{figure*}[ht!]
\centering\includegraphics[width=18cm]{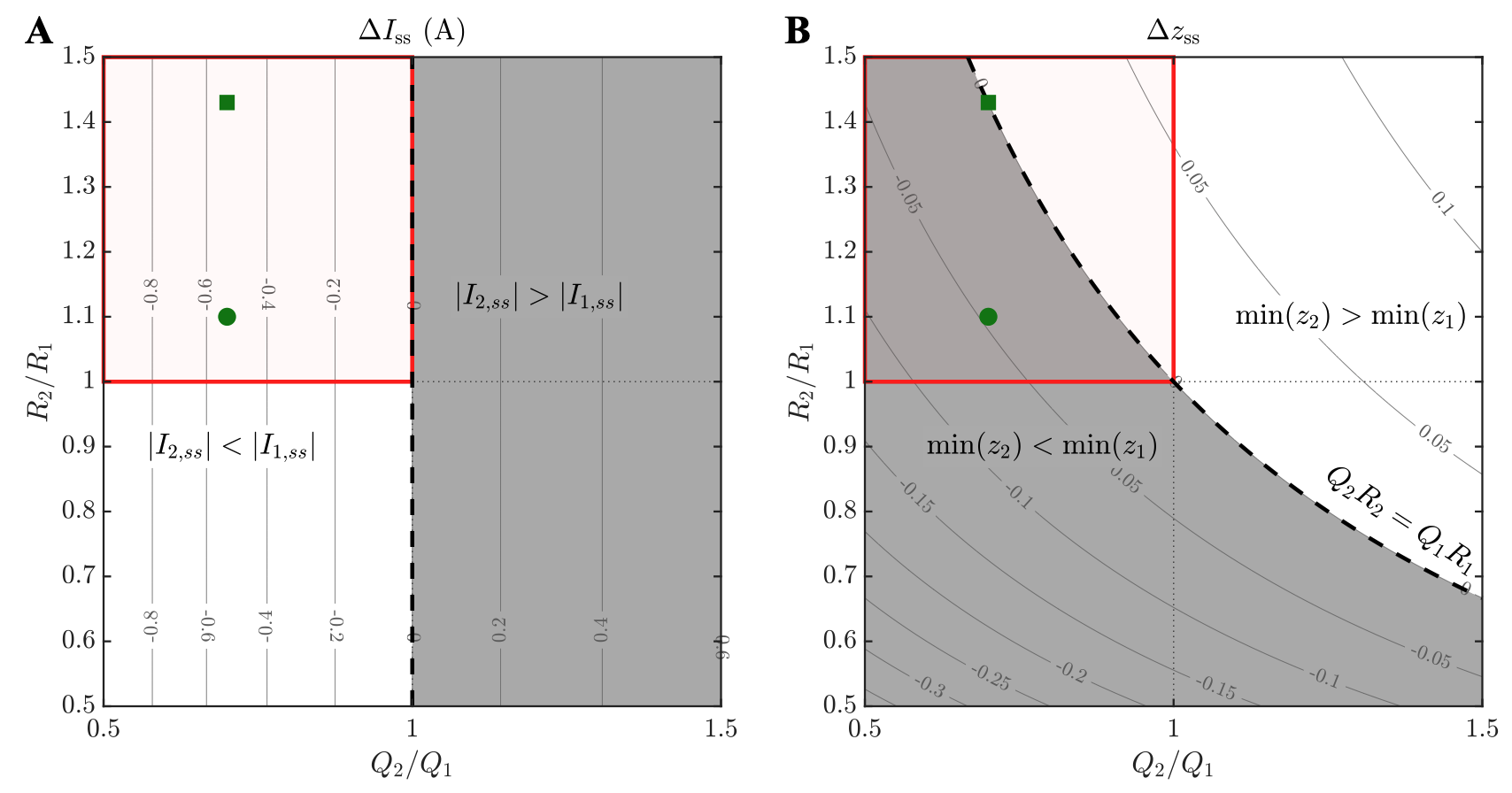}
\caption{Maps of current and SOC imbalances at steady-state as a function of capacity and resistance variability. (A) Steady-state current imbalance $\Delta I_{ss}$ during CC charge or discharge. (B) Steady-state SOC imbalance $\Delta z_{ss}$ during discharge. Red boxes show an aged cell (Cell 2) connected with a fresh cell (Cell 1), i.e. $Q_2 < Q_1$ and $R_2 > R_1$. Green markers correspond to the conditions shown in Figs. \ref{fig:affine-1} and \ref{fig:nonlinear-combined}. White regions indicate scenarios in which degradation is expected to converge (see Section \ref{sec:intra-to-inter}). Gray regions indicate scenarios in which degradation is expected to diverge. In Panel A, the convergence behavior is assumed to be driven by current magnitude differences (see Sections \ref{sec:current-imbalance-convergence}, \ref{sec:simulating-convergence}). In Panel B, the convergence/divergence behavior is assumed to be driven by differences in depths of discharge or cell utilization (see Section \ref{sec:soc-imbalance-convergence}).}
\label{fig:affine-map-combined}
\end{figure*}

\subsection{Current Imbalance Can Lead to Degradation Convergence}
\label{sec:current-imbalance-convergence}

\revb{We now describe a set of assumptions under which current imbalance can lead to degradation convergence.} Suppose that, for each cell $i$, the incremental capacity lost over a single cycle is proportional to the maximum current experienced during that cycle. This assumption approximates the lithium consumption process during SEI growth or lithium plating, both of which are accelerated at higher current densities due to Butler-Volmer kinetics \cite{Ning2006-hb, Yang2017-uf, Pinson2012-wt}. 

Using the incremental capacity loss model developed in Section \ref{sec:incremental}, we can represent the effect of current on the degradation rate as a modification to the reaction rate parameter:
\begin{equation}
    \label{eq:rrr}
    r_{i,n} = \gamma_1\cdot I_{ss,i,n},
\end{equation}
where $\gamma_1$ is a proportionality constant and $I_{ss,i,n}$ is the steady-state current for cell $i$ and at cycle $n$. Here, the steady-state current provides a measure of the maximum current imbalance over the course of a cycle, assuming affine OCV-R model dynamics (Section \ref{sec:affine-properties}). Also note that (\ref{eq:rrr}) represents a simplification of the function $g$ from (\ref{eq:rin-general}) in which the only factor that influences the reaction rate is the steady-state current.

Degradation convergence requires that the aged cell (Cell 2) degrade more slowly than the fresh cell (Cell 1), or:
\begin{equation}
    r_{2,n} < r_{1,n}.
\end{equation} 
Expanding this inequality using (\ref{eq:rrr}) yields:
\begin{equation}
    \Delta I_{ss,n} < 0,
\end{equation}
where $\Delta I_{ss,n}$ is the steady-state current imbalance at cycle $n$ as previously defined in (\ref{eq:diss}). Substituting (\ref{eq:diss}) into this expression yields simply $Q_2 < Q_1$ which is satisfied by definition. The capacity degradation rate for the aged cell will therefore always be lower than that of the fresh cell. The capacity difference between the aged cell and the fresh cell will thus converge over \rev{the course of repeated} cycles.

Fig. \ref{fig:affine-map-combined}A \revb{further visualizes why, in the affine OCV-R system, current imbalance `favors' the aged cell while `penalizing' the fresh cell.} The red box highlights the region corresponding to when Cell 1 is a fresh cell and Cell 2 is an aged cell. The plot shows that the aged cell always experiences less current compared to the fresh cell at steady-state. This result is consistent with the condition simulated in Fig. \ref{fig:affine-1}, highlighted as a green circle. 

We note that Song et al. \cite{Song2021-dz} provided a similar analytical proof which also assumed affine OCV-R dynamics and a similar capacity loss model. This work showed that, with non-concave capacity degradation trajectories ($p<1$), cell-to-cell variability in capacity decreases over age, i.e. degradation trajectories converge.
 
\begin{figure*}[ht!]
\centering\includegraphics[width=\linewidth]{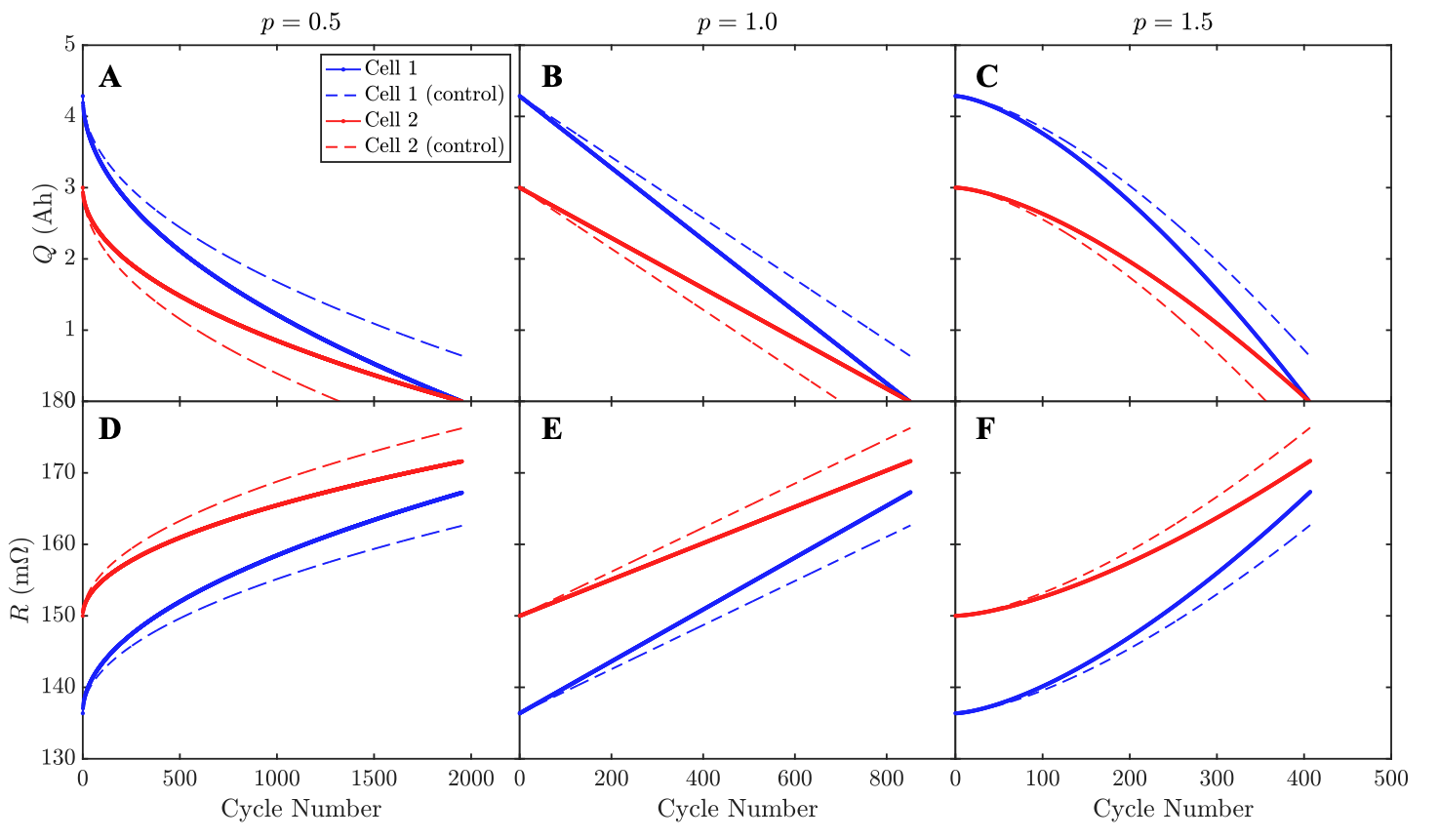}
\caption{Simulation of degradation convergence due to current imbalance. This result uses the incremental capacity degradation model to represent inter-cycle dynamics (Section \ref{sec:degradation}) and the affine OCV-R model to represent the intra-cycle dynamics (Section \ref{sec:affine}). In these simulations, steady-state current values (Eq. \ref{eq:rrr}) are assumed to drive cell capacity fade and resistance growth. Subpanels show simulations of capacity fade performed using different values of the SEI exponential growth factor $p$. A: self-limiting, B: linear, C: accelerating. (D-F) show the corresponding resistance growth simulations.}
\label{fig:degradation_convergence}
\end{figure*}

\subsection{Simulating Degradation Convergence Due to Current Imbalance}
\label{sec:simulating-convergence}

Degradation convergence induced by current imbalance can also be numerically demonstrated by simulating both the inter and intra-cycle dynamics according to the framework originally proposed in Fig. \ref{fig:inter-intra}. The simulation was initialized using the same cell parameters presented in Fig. \ref{fig:affine-1}. For each cycle, the steady-state current value was calculated based on Eqs. (\ref{eqn:I_2}) and (\ref{eqn:I_1}). The reaction rate constant for each cell was then updated according to (\ref{eq:rrr}). The incremental capacity loss was finally updated according to (\ref{eq:delta_l_simplified}). The process was repeated until the lowest cell capacity reached zero. The procedure used in the simulation is summarized in Algorithm \ref{alg:inter-intra}. Control cells were additionally simulated. For these cells, the current used to update the reaction rate was set to $I/2$ for all cycles.

\begin{algorithm}
\caption{Successive update scheme for 2 cells}
\label{alg:inter-intra}
\begin{algorithmic}
\Require $Q_{1,0}, Q_{2,0}, R_{1,0}, R_{2,0}$            \Comment{Initial Conditions}
\State $L_{1,0} \gets 0$                                 
\State $L_{2,0} \gets 0$                                 
\State $n \gets 1 $                                     
\While{($Q_{i,0} > Q_{\mathrm{min}})\forall i\in\{1,2\}$}
\For{each $i\in\{1,2\}$}                                 \Comment{Intra-Cycle Updates}
    \State $I_{i,n}(t) \gets f_1(t, Q_{1,n-1}, Q_{2,n-1},R_{1,n-1}, R_{2,n-1})$
    \State $z_{i,n}(t) \gets f_2(t, Q_{1,n-1}, Q_{2,n-1},R_{1,n-1}, R_{2,n-1})$
\EndFor
\For{each $i\in\{1,2\}$}                                    \Comment{Inter-Cycle Updates}
    \State $r_{i,n}(t) \gets g(I_{i,n}(t), z_{i,n}(t))$
    \State $\delta L_{i,n} \gets \left(\int_{t_{0,n}}^{t_{f,n}}r_{i,n}(\tau)^{1/p}d\tau + L_{i,n-1}^{1/p}\right)^p - L_{i,n-1}$
    \State $Q_{i,n} \gets Q_{i,n-1} - \delta L_{i,n}$
    \State $R_{i,n} \gets R_{i,n-1} + \lambda_1 \delta L_{i,n} + \lambda_2$
\EndFor
\State $n \gets n + 1$
\EndWhile
\end{algorithmic}
\end{algorithm}

Fig. \ref{fig:degradation_convergence} shows the simulation results under three values of $p$, representing three distinct degradation trajectories: self-limiting (A), linear (B), and accelerating (C). In all three cases, the aged cell (Cell 2), with lower initial capacity and higher initial resistance, lost capacity more slowly than the fresh cell (Cell 1). In fact, both cells reached zero capacity at exactly the same cycle number, irrespective of the value of $p$. The steady-state current values for both cells remained constant since the ratios $Q_1/\qtot$ and $Q_2/\qtot$ remained invariant even as $Q_1$ and $Q_2$ individually decreased. The current imbalance dynamics thus remained invariant over all cycles, with Cell 2 always seeing less current at steady-state than Cell 1.

Panels (D-F) show the corresponding resistance growth predictions. Resistance growth was calculated according to (\ref{eq:resistance}) with $(\lambda_1, \lambda_2)$ set to $(2\times10^{-6},0)$ for demonstration purposes. The results show that resistance growth trajectories also converge for all tested values of $p$.

\subsection{Degradation Convergence is not Guaranteed: Effect of SOC Imbalance on Depth of Discharge (DOD)}
\label{sec:soc-imbalance-convergence}

\rev{While the previous section showed that current imbalance can lead to degradation convergence, we now highlight another set of degradation assumptions that do not guarantee degradation convergence.}

Here, we assume that the cell degradation rate increases as the minimum SOC experienced by the cell at the end of each discharge is decreased. This assumption could be justified considering that a lower minimum SOC is equivalent to a higher depth of discharge (DOD). Higher DODs may amplify cathode particle cracking mechanisms, leading to higher capacity fade, higher resistance growth, or both \cite{Watanabe2014-fj, Li2018-ku, Gauthier2022-xt}. Note that, at the end of discharge, SOC re-balancing does not typically occur under practical applications which generally lack CV holds at the end of discharge cycles. It is thus possible that some cells in a parallel-connected group end discharge at lower SOCs compared to their neighbors. Under these assumptions, the reaction rate from Eq. \ref{eq:rrr} could take the following form:
\begin{equation}
    \label{eq:zzz}
    r_{i,n} = \frac{\gamma_2}{\mathrm{min}(z_{i,n}) + 1},
\end{equation}
where $\gamma_2>0$ is another proportionality constant and $\mathrm{min}(z_{i,n})$ is the minimum SOC experienced by cell $i$ at the end of discharge and for the $n$th cycle. The reaction rate here reaches a maximum value when $z_{i,n}$ approaches zero.

Figure \ref{fig:affine-map-combined}B shows how the steady-state SOC imbalance $\Delta z_{ss} \triangleq z_2 - z_1$ (Eq. \ref{eq:dzss}) can be used to predict which cell will end at a lower SOC. When $\Delta z_{ss} > 0$, Cell 2 (the aged cell) will end discharge with a higher SOC (i.e. lower DOD) and be degraded more slowly, leading to convergent degradation. We can thus interpret $\Delta z_{ss} > 0$ as the necessary condition for degradation convergence. However, according to Eq. (\ref{eq:dzss}), $\Delta z_{ss} > 0$ is only guaranteed if $Q_2R_2 > Q_1R_1$. Graphically, this condition corresponds to the white region in Figure \ref{fig:affine-map-combined}B. Recalling that the red box represents scenarios in which an aged cell (Cell 2) is connected with a fresh cell (Cell 1), we realize that degradation convergence is no longer guaranteed for all of these cases of interest. The green circle highlights one such case, wherein the aged cell ends discharge at a lower SOC (see Fig. \ref{fig:affine-1}), suggesting a higher DOD utilization and thus divergent degradation.

This simple example highlights that different degradation assumptions lead to different conclusions about the convergence and divergence of degradation trajectories. While the result from Sections \ref{sec:current-imbalance-convergence} and \ref{sec:simulating-convergence} suggested degradation convergence, these results were obtained assuming that current imbalance was the sole driver for cell degradation. When other factors such as SOC imbalance are added to the mix, results may differ.

\section{Experimental Verification}
\label{sec:experimental}

\reva{The utility of the modeling and analysis framework we propose in this work ultimately depends on its ability to predict real-world data. This section thus focuses on comparing model-based predictions to measured lab data.} Section \ref{sec:methods} outlines the experimental methods used to generate the lab data. Section \ref{sec:intra-cycle-validation} compares modeled versus measured intra-cycle dynamics. Section \ref{sec:inter-cycle-validation} discusses measured degradation convergence outcomes.

\subsection{Methods}
\label{sec:methods}

Two 2.5Ah lithium-ion pouch cells were built on a prototype battery manufacturing line. The cells both used graphite as the anode and NMC as the cathode. Each cell was individually pre-conditioned to different capacities and resistances by aging them via a 1C charge, 1C discharge cycling test protocol. Voltage limits during the cycling test were set to 3.0V to 4.2V. See Weng et al. \cite{Weng2021-qc, Weng2023-lj} for more details on the cell build process and experimental setup for the cycling tests.

Cell capacities and resistances were measured after pre-conditioning. Capacities were measured using a C/20 discharge from 4.2V to 3.0V. Resistances were calculated by taking the voltage difference between a C/20 charge curve and C/20 discharge curve, dividing this voltage difference by two times the input current, and averaging across all SOCs. The cell properties were measured to be ($Q_1, R_1$) = (2.11Ah, 201m$\Omega$), ($Q_2, R_2)$ = (1.83Ah, 250m$\Omega$).

\begin{figure}[ht!]
\centering\includegraphics[width=\linewidth]{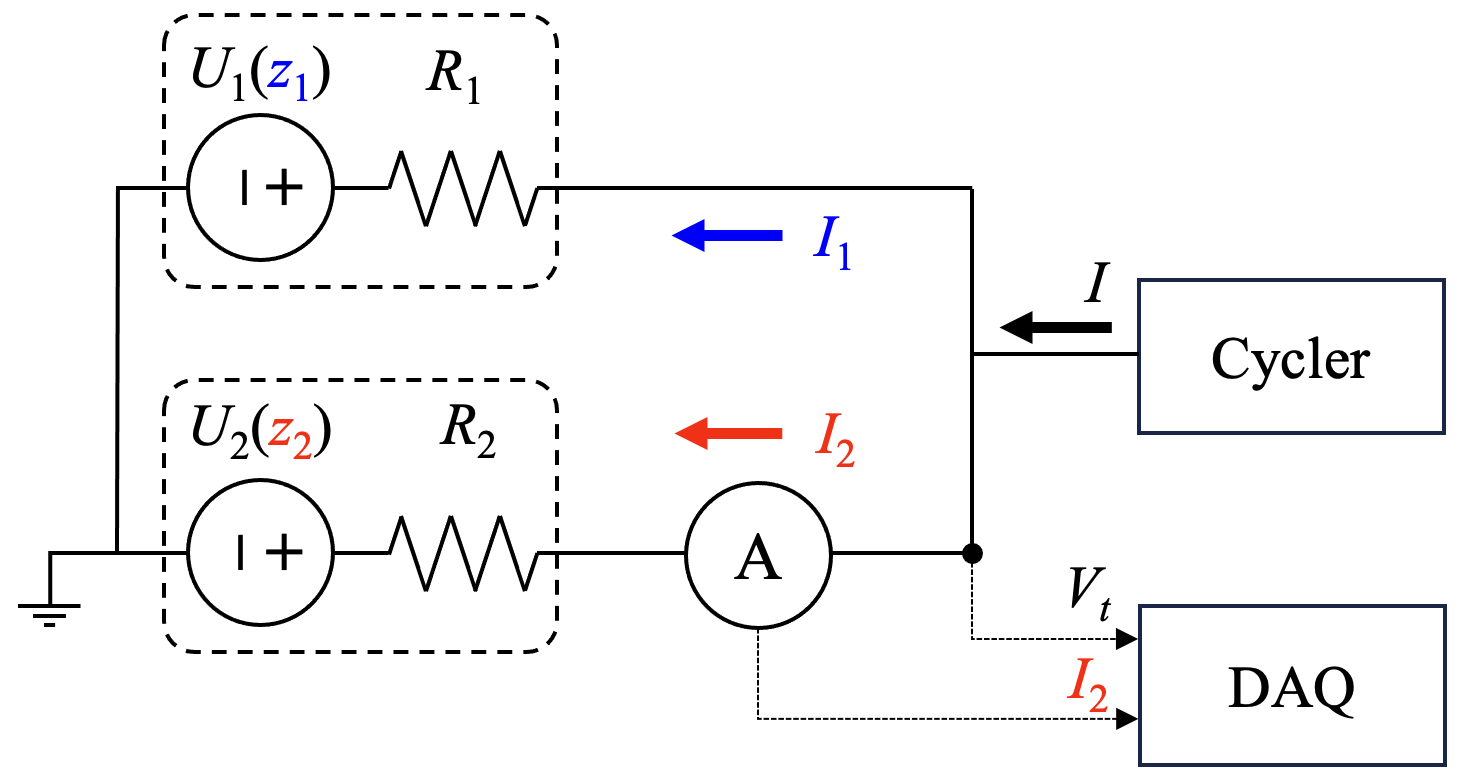}
\caption{Experimental setup for measuring branch currents in two parallel-connected cells.}
\label{fig:setup}
\end{figure}

After pre-conditioning, cells were connected in a parallel arrangement and cycled using a commercial cycler (Arbin BT2000), according to Fig. \ref{fig:setup}. The cycling profile consisted of 1C CCCV charge and 1C CC discharge between 3.0V and 4.2V. The branch current for Cell 2 was measured using a commercial Hall effect sensor (Allegro ACS70331) and logged using LabVIEW. The branch current for Cell 1 was calculated from the difference between the total input current and the branch current for Cell 2. Resistances of the wires and the Hall effect sensor were estimated to be less than 5m$\Omega$ in each branch and were thus ignored. 

The parallel-connected system was cycled for 240 equivalent full cycles, then stopped. The system was then charged and discharged at different rates, including C/4 and C/10, to characterize the intra-cycle dynamics for different input currents. Finally, the cells were disconnected from the parallel connection to re-measure individual cell capacities and resistances using the same process described above.

\subsection{Intra-Cycle Dynamics}
\label{sec:intra-cycle-validation}

To compare the model-predicted intra-cycle dynamics against experiment, the model was initialized using measured values for $(Q_1,R_1,Q_2,R_2)$. The nonlinear OCV functions $U_1$ and $U_2$ were parameterized by taking the average of the C/20 charge and C/20 discharge voltage curves. The SOC for each cell was initialized to match the measured terminal voltage preceding the charge cycle (ca. 0.3\%). The model was simulated using the method described in Section \ref{sec:nonlinear-simulations}.

\begin{figure*}[ht!]
\centering\includegraphics[width=\linewidth]{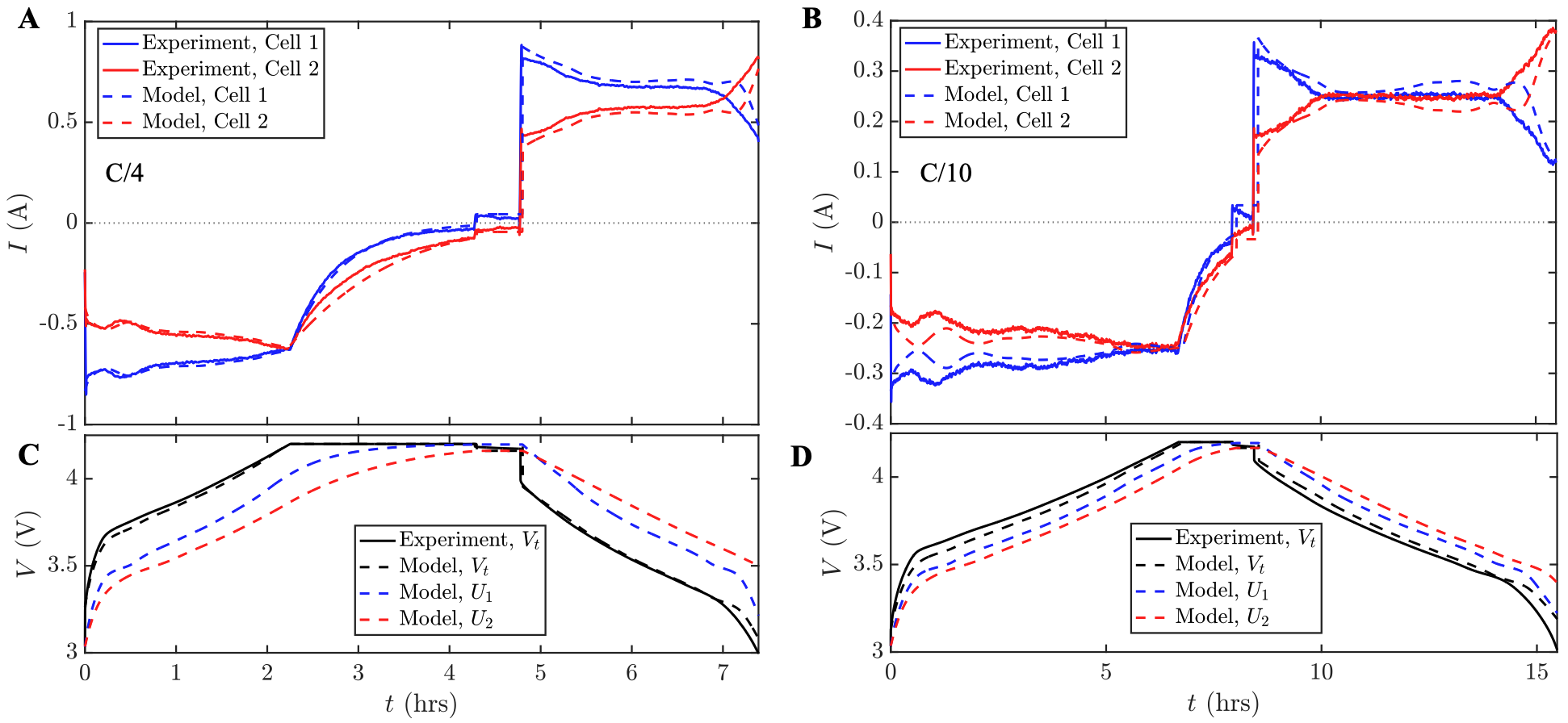}
\caption{Model versus experimental comparison of intra-cycle dynamics for two cells with mismatched capacities and resistances. ($Q_1,R_1$) = (1.83Ah, 549m$\Omega$) and ($Q_2, R_2$) = (1.93Ah, 277m$\Omega$). The experimental profile consists of CCCV charging, resting for 30 minutes, followed by CC discharging. (A) C/4 charge and discharge. (B) C/10 charge and discharge. (C,D) The corresponding voltage profiles. The CV hold cut-off condition is set to C/30 for all cases.}
\label{fig:validation}
\end{figure*}

Fig. \ref{fig:validation} compares the modeled versus measured intra-cycle dynamics at two different C-rates: C/4 (Panels A,C) and C/10 (Panels B,D). At C/4, the model-predicted current imbalances and terminal voltages show good agreement with the data. The model captured inflections in the current imbalance and correctly predicted the durations of both the CC and CV phases. This result is surprising considering the simplicity of the model which omits RC circuit elements and assumes a constant value for resistance. At C/10, the model-predicted current imbalances qualitatively match the experimental results, though some model mismatches are evident. Specifically, during the CC charge, the model under-predicted the magnitude of the current imbalance, and during the CC discharge, the model over-predicted the current imbalance at mid-SOCs. The model also under-predicted the measured terminal voltage on both charge and discharge. The cause of the model mismatches at low currents is unclear and suggests the need for future investigations.

\subsection{Inter-Cycle Dynamics}
\label{sec:inter-cycle-validation}

Table \ref{tbl:inter-cycle-validation} compares the measured cell capacities and resistances before and after the parallel cycling test. The measured results suggest that the capacities are converging, with $Q_2/Q_1$ increasing from 0.938 before cycling to 0.948 after cycling. However, the resistances are not converging but diverging: $R_2/R_1$ has increased from 1.59 to 1.99. This result thus highlights a scenario in which initial variability in cell properties has led to a convergence in the individual cell capacities but divergence in the individual cell resistances. Degradation convergence is thus not a universal guarantee and depends on the degradation metric (capacity or resistance). Degradation convergence is also likely strongly influenced by the underlying degradation assumptions. For example, Section \ref{sec:soc-imbalance-convergence} explored how a degradation assumption based solely on current imbalance led to convergent degradation, but a degradation assumption based solely on SOC imbalance could lead to either convergent or divergent degradation depending on the values of the cell capacities and resistances.

\begin{table}[ht!]
\caption{Comparison of experimentally-measured cell capacities and resistances before and after parallel cycling.}
\begin{center}
\label{tbl:inter-cycle-validation}
\begin{tabular}{c c c}
\hline
 & Before Cycling & After Cycling \\
\hline
$Q_2$ & 1.98 Ah & 1.83 Ah \\
$Q_1$ & 2.11 Ah & 1.93 Ah \\
$Q_2/Q_1$ & 0.938 & 0.948 \\
$R_2$ & 321 m$\Omega$ & 550 m$\Omega$ \\
$R_1$ & 201 m$\Omega$ & 277 m$\Omega$ \\
$R_2/R_1$ & 1.59 & 1.99 \\
\end{tabular}
\end{center}
\end{table}

\section{Future Work Recommendations}
\label{sec:future}

\revb{The modeling framework derived in this work lends itself to mathematical analysis, enabling more rigorous treatments of degradation convergence or divergence for future studies. We envision that this framework can be expanded in the future to support analyzing parallel-connected systems in a more generalized context. This section highlights several areas for future exploration}:

\textbf{a. Arbitrary number of parallel-connected cells.} This work presented only two cells connected parallel to highlight the effect of initial cell variability on an elementary system. Since only two cells were considered, the presented results may not generalize to the case of $n>2$ cells connected in parallel. However, our analysis can be extended to the general case of $n$-cells by leveraging ideas presented by Song et al. \cite{Song2021-dz, Song2022-fg} and Drummond et al. \cite{Drummond2021-pp} to resolve the algebraic constraint for the branch currents and describing the solutions in state-space form.

\textbf{b. Thermal analysis.} Considerations for thermal imbalances (i.e. due to battery pack design or differences in individual cell heating rates) were also omitted in this work but will be important to consider for future work. To understand thermal effects, our proposed modeling approach can be combined with a cell thermal model such as those presented by Zhang et al., \cite{Zhang2020-dg}, Song et al. \cite{Song2021-dz, Song2022-fg}, Reniers et al. \cite{Reniers2023-bg} and Hosseinzadeh et al. \cite{Hosseinzadeh2021-ma}. Thermal model predictions should also be compared against recent experimental work including those from Fill et al. \cite{Fill2021-je}, Paarmann et al. \cite{Paarmann2021-yt} and Naylor-Marlow et al. \cite{Naylor-Marlow2023-gd}.

\textbf{b. Interconnect resistances.} Interconnect resistances also play an important role in determining current imbalances in real battery packs. This work ignored interconnect resistances to simplify the analysis. Other authors such as Reniers et al. \cite{Reniers2023-bg} and Hosseinzadeh et al. \cite{Hosseinzadeh2021-ma} did consider interconnect resistances in their work. Ideas presented in these works can be incorporated in the future to understand the relative contribution of cell versus non-cell components on the overall system current imbalance.

\textbf{c. Degradation mechanisms.} Our experimental work showed that degradation convergence is not guaranteed in parallel-connected systems. Specifically, while capacity fade trajectories appeared to be converging, resistance growth trajectories appeared to be diverging. The divergent resistance growth trajectory was not predicted by our modeling framework which assumed a single degradation mechanism: SEI growth. In reality, degradation pathways in real lithium-ion devices are multi-faceted \cite{Birkl2017-yq, Han2019-vh, Woody2020-ah, Edge2021-fy, Jiang2021-xg, Attia2022-jm}, requiring more sophisticated degradation models \cite{OKane2022-fj, Brosa_Planella2022-aq, Brosa_Planella2023-ae} and parameterization methods \cite{Chen2020-og, Wang2022-tl} to fully capture. Future work should thus explore how additional degradation assumptions could influence the convergence behavior in parallel-connected systems using a combination of simulation-based and analytical approaches.

\section{Conclusion}

This work proposed a framework to model coupled degradation phenomena in parallel-connected battery systems. \revb{We first developed an} incremental capacity loss scheme based on SEI growth dynamics which was used to compute the cycle-to-cycle (i.e. inter-cycle) evolution of the capacity loss and resistance growth of individual cells connected in parallel. \revb{We then developed} a set of closed-form, analytical solutions to the affine OCV-R model which described the dynamics of current and SOC imbalance within a single cycle (i.e. intra-cycle). The two components were coupled by an SEI reaction rate term which was made a function of the intra-cycle dynamical variables such as the current and SOC for each individual cell.

The degradation analysis demonstrated in this work \revb{assumed a single degradation mode, SEI growth. With this assumption, we demonstrated that current imbalance can lead to convergent degradation trajectories. However, we further presented experimental evidence that degradation convergence may not be guaranteed, and that convergent capacity fade trajectories does not imply convergent resistance growth trajectories. Understanding this phenomenon will require the inclusion of additional degradation mechanisms.}

\revb{The modeling and analysis framework developed here can be extended to further study general degradation phenomena in parallel-connected battery systems.} Such developments remain necessary to fully answer the unavoidable yet essential questions in battery cell manufacturing (``how much manufacturing variability is too much?'') and pack re-purposing (``how much dissimilarity can be tolerated when repairing old packs using fresher counterparts?'').

\begin{nomenclature}

\noindent Indices

\begin{itemize}
    \item $i$ : cell number
    \item $n$ : cycle number
\end{itemize}

\noindent OCV-R Model

\begin{itemize}
    \item $z$ : cell state of charge 
    \item $R$ : cell internal resistance [Ohms]
    \item $Q$ : cell capacity [Amp-seconds]
    \item $I$ : cell branch current [Amps]
    \item $U$ : cell open-circuit voltage [Volts]
    \item $\alpha$ : affine OCV slope parameter [Volts]
    \item $\beta$ : affine OCV minimum voltage parameter [Volts]
    \item $\kappa$ : input sensitivity factor [Amps$^{-1}$]
    \item $\tau$ : time constant [Seconds]
\end{itemize}

\noindent Acronyms
\begin{itemize}
    \item CC : constant current
    \item CV : constant voltage
    \item Gr : graphite
    \item LFP : lithium iron phosphate
    \item NMC : nickel manganese cobalt
    \item OCV : open circuit voltage
    \item SEI : solid electrolyte interphase
    \item SOC : state of charge
\end{itemize}

\noindent Degradation Model
\begin{itemize}
    \item $L$ : total capacity lost 
    \item $G$ : total resistance growth
    \item $r$ : reaction rate constant
    \item $\gamma$: reaction rate proportionality constant
    \item $p$ : reaction rate exponential factor
    \item $\lambda$ : capacity to resistance proportionality constant
\end{itemize}

\end{nomenclature}

%%%%%%%%%%%%%%%%%%%%%%%%%%%%%%%%%%%%%%%%%%%%%%%%%%%%%%%%%%%%%%%%%%%%%%
% The bibliography is stored in an external database file
% in the BibTeX format (file_name.bib).  The bibliography is
% created by the following command and it will appear in this
% position in the document. You may, of course, create your
% own bibliography by using thebibliography environment as in
%
% \begin{thebibliography}{12}
% ...
% \bibitem{itemreference} D. E. Knudsen.
% {\em 1966 World Bnus Almanac.}
% {Permafrost Press, Novosibirsk.}
% ...
% \end{thebibliography}

% Here's where you specify the bibliography style file.
% The full file name for the bibliography style file 
% used for an ASME paper is asmems4.bst.
\bibliographystyle{asmems4}

% Here's where you specify the bibliography database file.
% The full file name of the bibliography database for this
% article is asme2e.bib. The name for your database is up
% to you.
\bibliography{main}

\begin{thebibliography}{10}

\bibitem{Baumhofer2014-xc}
Baumh{\"o}fer, T., Br{\"u}hl, M., Rothgang, S., and Sauer, D.~U., 2014.
\newblock ``Production caused variation in capacity aging trend and correlation to initial cell performance''.
\newblock {\em J. Power Sources, {\bf 247}}, pp.~332--338.

\bibitem{Schindler2021-pn}
Schindler, M., Sturm, J., Ludwig, S., Schmitt, J., and Jossen, A., 2021.
\newblock ``Evolution of initial cell-to-cell variations during a three-year production cycle''.
\newblock {\em eTransportation, {\bf 8}}, p.~100102.

\bibitem{Wildfeuer2021-xs}
Wildfeuer, L., and Lienkamp, M., 2021.
\newblock ``Quantifiability of inherent cell-to-cell variations of commercial lithium-ion batteries''.
\newblock {\em eTransportation, {\bf 9}}, Aug., p.~100129.

\bibitem{Kenney2012-hl}
Kenney, B., Darcovich, K., MacNeil, D.~D., and Davidson, I.~J., 2012.
\newblock ``Modelling the impact of variations in electrode manufacturing on lithium-ion battery modules''.
\newblock {\em J. Power Sources, {\bf 213}}, pp.~391--401.

\bibitem{Schmidt2020-mi}
Schmidt, O., Thomitzek, M., R{\"o}der, F., Thiede, S., Herrmann, C., and Krewer, U., 2020.
\newblock ``Modeling the impact of manufacturing uncertainties on {Lithium-Ion} batteries''.
\newblock {\em J. Electrochem. Soc., {\bf 167}}(6), p.~060501.

\bibitem{Weng2023-lj}
Weng, A., Siegel, J.~B., and Stefanopoulou, A., 2023.
\newblock ``Differential voltage analysis for battery manufacturing process control''.
\newblock {\em Frontiers in Energy Research, {\bf 11}}.

\bibitem{Harper2019-uq}
Harper, G., Sommerville, R., Kendrick, E., Driscoll, L., Slater, P., Stolkin, R., Walton, A., Christensen, P., Heidrich, O., Lambert, S., Abbott, A., Ryder, K., Gaines, L., and Anderson, P., 2019.
\newblock ``Recycling lithium-ion batteries from electric vehicles''.
\newblock {\em Nature, {\bf 575}}(7781), pp.~75--86.

\bibitem{Chen2019-sb}
Chen, M., Ma, X., Chen, B., Arsenault, R., Karlson, P., Simon, N., and Wang, Y., 2019.
\newblock ``Recycling {End-of-Life} electric vehicle {Lithium-Ion} batteries''.
\newblock {\em Joule, {\bf 3}}(11), pp.~2622--2646.

\bibitem{Lai2021-xa}
Lai, X., Huang, Y., Deng, C., Gu, H., Han, X., Zheng, Y., and Ouyang, M., 2021.
\newblock ``Sorting, regrouping, and echelon utilization of the large-scale retired lithium batteries: A critical review''.
\newblock {\em Renewable Sustainable Energy Rev., {\bf 146}}(April), p.~111162.

\bibitem{Wang2021-uz}
Wang, X., Fang, Q., Dai, H., Chen, Q., and Wei, X., 2021.
\newblock ``Investigation on cell performance and inconsistency evolution of series and parallel {Lithium-Ion} battery modules''.
\newblock {\em Energy Technology, {\bf 9}}(7), pp.~1--10.

\bibitem{Zilberman2020-uw}
Zilberman, I., Schmitt, J., Ludwig, S., Naumann, M., and Jossen, A., 2020.
\newblock ``Simulation of voltage imbalance in large lithium-ion battery packs influenced by cell-to-cell variations and balancing systems''.
\newblock {\em Journal of Energy Storage, {\bf 32}}(April), p.~101828.

\bibitem{Rasheed2020-ly}
Rasheed, M., Kamel, M., Wang, H., Zane, R., and Smith, K., 2020.
\newblock ``Investigation of active life balancing to recondition li-ion battery packs for 2ndlife''.
\newblock {\em 2020 IEEE 21st Workshop on Control and Modeling for Power Electronics, COMPEL 2020}.

\bibitem{Feng2019-zq}
Feng, X., Xu, C., He, X., Wang, L., Gao, S., and Ouyang, M., 2019.
\newblock ``A graphical model for evaluating the status of series-connected lithium-ion battery pack''.
\newblock {\em Int. J. Energy Res., {\bf 43}}(2), pp.~749--766.

\bibitem{Chen2023-la}
Chen, J., Ouyang, Q., and Wang, Z., 2023.
\newblock ``Overview of cell equalization systems''.
\newblock In {\em Equalization Control for Lithium-ion Batteries}, J.~Chen, Q.~Ouyang, and Z.~Wang, eds. Springer Nature Singapore, Singapore, pp.~13--28.

\bibitem{Lin2020-tb}
Lin, X., Perez, H.~E., Siegel, J.~B., and Stefanopoulou, A.~G., 2020.
\newblock ``Robust estimation of battery system temperature distribution under sparse sensing and uncertainty''.
\newblock {\em IEEE Trans. Control Syst. Technol.}

\bibitem{Gong2015-mb}
Gong, X., Xiong, R., and Mi, C.~C., 2015.
\newblock ``Study of the characteristics of battery packs in electric vehicles with {Parallel-Connected} {Lithium-Ion} battery cells''.
\newblock {\em IEEE Trans. Ind. Appl., {\bf 51}}(2), Mar., pp.~1872--1879.

\bibitem{Brand2016-hc}
Brand, M.~J., Hofmann, M.~H., Steinhardt, M., Schuster, S.~F., and Jossen, A., 2016.
\newblock ``Current distribution within parallel-connected battery cells''.
\newblock {\em J. Power Sources, {\bf 334}}, Dec., pp.~202--212.

\bibitem{Luca2021-up}
Luca, R., Whiteley, M., Neville, T., Tranter, T., Weaving, J., Marco, J., Shearing, P.~R., and Brett, D. J.~L., 2021.
\newblock ``Current imbalance in parallel battery strings measured using a {Hall-Effect} sensor array''.
\newblock {\em Energy Technology, {\bf 9}}(4), pp.~1--11.

\bibitem{Bruen2016-vr}
Bruen, T., and Marco, J., 2016.
\newblock ``Modelling and experimental evaluation of parallel connected lithium ion cells for an electric vehicle battery system''.
\newblock {\em J. Power Sources, {\bf 310}}, pp.~91--101.

\bibitem{Song2021-dz}
Song, Z., Yang, X.-G., Yang, N., Delgado, F.~P., Hofmann, H., and Sun, J., 2021.
\newblock ``A study of cell-to-cell variation of capacity in parallel-connected lithium-ion battery cells''.
\newblock {\em eTransportation, {\bf 7}}, Feb., p.~100091.

\bibitem{Song2022-fg}
Song, Z., Yang, N., Lin, X., Delgado, F.~P., Hofmann, H., and Sun, J., 2022.
\newblock ``Progression of cell-to-cell variation within battery modules under different cooling structures''.
\newblock {\em Appl. Energy}.

\bibitem{Reniers2023-bg}
Reniers, J.~M., and Howey, D.~A., 2023.
\newblock ``Digital twin of a {MWh-scale} grid battery system for efficiency and degradation analysis''.
\newblock {\em Appl. Energy, {\bf 336}}, Apr., p.~120774.

\bibitem{Liu2019-yf}
Liu, X., Ai, W., Naylor~Marlow, M., Patel, Y., and Wu, B., 2019.
\newblock ``The effect of cell-to-cell variations and thermal gradients on the performance and degradation of lithium-ion battery packs''.
\newblock {\em Appl. Energy, {\bf 248}}(April), pp.~489--499.

\bibitem{Paarmann2021-zq}
Paarmann, S., Cloos, L., Technau, J., and Wetzel, T., 2021.
\newblock ``Measurement of the temperature influence on the current distribution in lithium‐ion batteries''.
\newblock {\em Energy Technol., {\bf 9}}(6), June, p.~2000862.

\bibitem{Hofmann2018-gg}
Hofmann, M.~H., Czyrka, K., Brand, M.~J., Steinhardt, M., Noel, A., Spingler, F.~B., and Jossen, A., 2018.
\newblock ``Dynamics of current distribution within battery cells connected in parallel''.
\newblock {\em Journal of Energy Storage, {\bf 20}}, Dec., pp.~120--133.

\bibitem{Zhang2020-ci}
Zhang, D., Couto, L.~D., Benjamin, S., Zeng, W., Coutinho, D.~F., and Moura, S.~J., 2020.
\newblock ``State of charge estimation of parallel connected battery cells via descriptor system theory''.
\newblock In 2020 American Control Conference (ACC), pp.~2207--2212.

\bibitem{Zhang2020-dg}
Zhang, D., Couto, L.~D., Gill, P., Benjamin, S., Zeng, W., and Moura, S.~J., 2020.
\newblock ``Interval observer for {SOC} estimation in {Parallel-Connected} lithium-ion batteries''.
\newblock {\em Proc. Am. Control Conf., {\bf 2020-July}}, pp.~1149--1154.

\bibitem{Drummond2021-pp}
Drummond, R., Couto, L.~D., and {others}, 2021.
\newblock ``Resolving kirchhoff's laws for parallel li-ion battery pack state-estimators''.
\newblock {\em IEEE transactions on}.

\bibitem{Li2022-kf}
Li, Z., Zuo, A., Mo, Z., Lin, M., Wang, C., Zhang, J., Hofmann, M.~H., and Jossen, A., 2022.
\newblock ``Demonstrating stability within parallel connection as a basis for building large-scale battery systems''.
\newblock {\em Cell Reports Physical Science}, Nov., p.~101154.

\bibitem{Fill2021-je}
Fill, A., and Peter~Birke, K., 2021.
\newblock ``Influences of cell to cell variances and the battery design on thermal and electrical imbalances among parallel {Lithium-Ion} cells''.
\newblock In Proceedings of the {IEEE} International Conference on Industrial Technology, Vol.~2021-March, Institute of Electrical and Electronics Engineers Inc., pp.~391--396.

\bibitem{Fill2018-dm}
Fill, A., Koch, S., Pott, A., and Birke, K.-P., 2018.
\newblock ``Current distribution of parallel-connected cells in dependence of cell resistance, capacity and number of parallel cells''.
\newblock {\em J. Power Sources, {\bf 407}}, Dec., pp.~147--152.

\bibitem{Chen2021-ax}
Chen, L., Zhang, M., Ding, Y., Wu, S., Li, Y., Liang, G., Li, H., and Pan, H., 2021.
\newblock ``Estimation the internal resistance of lithium-ion-battery using a multi-factor dynamic internal resistance model with an error compensation strategy''.
\newblock {\em Energy Reports, {\bf 7}}, Nov., pp.~3050--3059.

\bibitem{Weng2021-qc}
Weng, A., Mohtat, P., Attia, P.~M., Sulzer, V., Lee, S., Less, G., and Stefanopoulou, A.~G., 2021.
\newblock ``Predicting the impact of formation protocols on battery lifetime immediately after manufacturing''.
\newblock {\em Joule}, pp.~1--22.

\bibitem{Guo2010-av}
Guo, M., Sikha, G., and White, R.~E., 2010.
\newblock ``{Single-Particle} model for a {Lithium-Ion} cell: Thermal behavior''.
\newblock {\em J. Electrochem. Soc., {\bf 158}}(2), Dec., p.~A122.

\bibitem{Moura2017-lg}
Moura, S.~J., Argomedo, F.~B., Klein, R., Mirtabatabaei, A., and Krstic, M., 2017.
\newblock ``Battery state estimation for a single particle model with electrolyte dynamics''.
\newblock {\em IEEE Trans. Control Syst. Technol., {\bf 25}}(2), Mar., pp.~453--468.

\bibitem{Prada2013-zz}
Prada, E., Di~Domenico, D., Creff, Y., Bernard, J., Sauvant-Moynot, V., and Huet, F., 2013.
\newblock ``A simplified electrochemical and thermal aging model of {LiFePO} 4 -graphite li-ion batteries: Power and capacity fade simulations''.
\newblock {\em J. Electrochem. Soc., {\bf 160}}(4), pp.~A616--A628.

\bibitem{Chen2020-tg}
Chen, C.-H., Planella, F.~B., O'Regan, K., Gastol, D., Dhammika~Widanage, W., and Kendrick, E., 2020.
\newblock ``Development of experimental techniques for parameterization of multi-scale lithium-ion battery models''.
\newblock {\em J. Electrochem. Soc., {\bf 167}}(8), May, p.~080534.

\bibitem{Khalil2002}
Khalil, H.~K., 2002.
\newblock {\em Nonlinear Systems, 3rd Edition}.
\newblock Prentice Hall.

\bibitem{Haddad2008}
Haddad, W.~M., and Chellaboina, V., 2008.
\newblock {\em Nonlinear Dynamical Systems and Control: A Lyapunov-Based Approach}.
\newblock Princeton University Press.

\bibitem{Smith2021-qv}
Smith, K., Gasper, P., Colclasure, A.~M., Shimonishi, Y., and Yoshida, S., 2021.
\newblock ``{Lithium-Ion} battery life model with electrode cracking and {Early-Life} break-in processes''.
\newblock {\em J. Electrochem. Soc., {\bf 168}}(10), Oct., p.~100530.

\bibitem{OKane2022-fj}
O'Kane, S. E.~J., Ai, W., Madabattula, G., Alonso-Alvarez, D., Timms, R., Sulzer, V., Edge, J.~S., Wu, B., Offer, G.~J., and Marinescu, M., 2022.
\newblock ``Lithium-ion battery degradation: how to model it''.
\newblock {\em Phys. Chem. Chem. Phys., {\bf 24}}(13), Mar., pp.~7909--7922.

\bibitem{Birkl2017-yq}
Birkl, C.~R., Roberts, M.~R., McTurk, E., Bruce, P.~G., and Howey, D.~A., 2017.
\newblock ``Degradation diagnostics for lithium ion cells''.
\newblock {\em J. Power Sources, {\bf 341}}, Feb., pp.~373--386.

\bibitem{Pinson2012-wt}
Pinson, M.~B., and Bazant, M.~Z., 2012.
\newblock ``Theory of {SEI} formation in rechargeable batteries: Capacity fade, accelerated aging and lifetime prediction''.
\newblock {\em Pinson, M. B., and M. Z. Bazant. ``Theory of SEI Formation in Rechargeable Batteries: Capacity Fade, Accelerated Aging and Lifetime Prediction.'' Journal of the Electrochemical Society, {\bf 16}}, Dec., p.~0.2.

\bibitem{Karger2022-kt}
Karger, A., Wildfeuer, L., Ayg{\"u}l, D., Maheshwari, A., Singer, J.~P., and Jossen, A., 2022.
\newblock ``Modeling capacity fade of lithium-ion batteries during dynamic cycling considering path dependence''.
\newblock {\em Journal of Energy Storage, {\bf 52}}, Aug., p.~104718.

\bibitem{Yang2017-uf}
Yang, X.~G., Leng, Y., Zhang, G., Ge, S., and Wang, C.~Y., 2017.
\newblock ``Modeling of lithium plating induced aging of lithium-ion batteries: Transition from linear to nonlinear aging''.
\newblock {\em J. Power Sources, {\bf 360}}, pp.~28--40.

\bibitem{Abraham2005-fd}
Abraham, D.~P., 2005.
\newblock Diagnostic examination of generation 2 {Lithium-Ion} cells and assessment of performance degradation mechanisms prepared by chemical engineering division.
\newblock Tech. rep., Argonne National Laboratory.

\bibitem{Ning2006-hb}
Ning, G., White, R.~E., and Popov, B.~N., 2006.
\newblock ``A generalized cycle life model of rechargeable li-ion batteries''.
\newblock {\em Electrochim. Acta, {\bf 51}}(10), pp.~2012--2022.

\bibitem{Watanabe2014-fj}
Watanabe, S., Kinoshita, M., Hosokawa, T., Morigaki, K., and Nakura, K., 2014.
\newblock ``Capacity fade of {LiAlyNi1-x-yCoxO} 2 cathode for lithium-ion batteries during accelerated calendar and cycle life tests (surface analysis of {LiAlyNi1-x-yCo} xo2 cathode after cycle tests in restricted depth of discharge ranges)''.
\newblock {\em J. Power Sources, {\bf 258}}, pp.~210--217.

\bibitem{Li2018-ku}
Li, J., Harlow, J., Stakheiko, N., Zhang, N., Paulsen, J., and Dahn, J., 2018.
\newblock ``Dependence of cell failure on {Cut-Off} voltage ranges and observation of kinetic hindrance in {LiNi} 0.8 co 0.15 al 0.05 {O} 2''.
\newblock {\em J. Electrochem. Soc., {\bf 165}}(11), pp.~A2682--A2695.

\bibitem{Gauthier2022-xt}
Gauthier, R., Luscombe, A., Bond, T., Bauer, M., Johnson, M., Harlow, J., Louli, A., and Dahn, J.~R., 2022.
\newblock ``How do depth of discharge, c-rate and calendar age affect capacity retention, impedance growth, the electrodes, and the electrolyte in {Li-Ion} cells?''.
\newblock {\em J. Electrochem. Soc.}, Jan.

\bibitem{Hosseinzadeh2021-ma}
Hosseinzadeh, E., Arias, S., Krishna, M., Worwood, D., Barai, A., Widanalage, D., and Marco, J., 2021.
\newblock ``Quantifying cell-to-cell variations of a parallel battery module for different pack configurations''.
\newblock {\em Appl. Energy, {\bf 282}}(PA), p.~115859.

\bibitem{Paarmann2021-yt}
Paarmann, S., Cloos, L., Technau, J., and Wetzel, T., 2021.
\newblock ``Measurement of the temperature influence on the current distribution in lithium‐ion batteries''.
\newblock {\em Energy Technol., {\bf 9}}(6), June, p.~2000862.

\bibitem{Naylor-Marlow2023-gd}
Naylor-Marlow, M., Chen, J., and Wu, B., 2023.
\newblock ``Battery pack degradation-understanding aging in parallel-connected lithium-ion batteries under thermal gradients''.

\bibitem{Han2019-vh}
Han, X., Lu, L., Zheng, Y., Feng, X., Li, Z., Li, J., and Ouyang, M., 2019.
\newblock ``A review on the key issues of the lithium ion battery degradation among the whole life cycle''.
\newblock {\em eTransportation, {\bf 1}}(August), p.~100005.

\bibitem{Woody2020-ah}
Woody, M., Arbabzadeh, M., Lewis, G.~M., Keoleian, G.~A., and Stefanopoulou, A., 2020.
\newblock ``Strategies to limit degradation and maximize li-ion battery service lifetime - critical review and guidance for stakeholders''.
\newblock {\em Journal of Energy Storage, {\bf 28}}, Apr., p.~101231.

\bibitem{Edge2021-fy}
Edge, J.~S., O'Kane, S., Prosser, R., Kirkaldy, N.~D., Patel, A.~N., Hales, A., Ghosh, A., Ai, W., Chen, J., Yang, J., Li, S., Pang, M.-C., Bravo~Diaz, L., Tomaszewska, A., Marzook, M.~W., Radhakrishnan, K.~N., Wang, H., Patel, Y., Wu, B., and Offer, G.~J., 2021.
\newblock ``Lithium ion battery degradation: what you need to know''.
\newblock {\em Phys. Chem. Chem. Phys., {\bf 23}}(14), Apr., pp.~8200--8221.

\bibitem{Jiang2021-xg}
Jiang, M., Danilov, D.~L., Eichel, R.-A., and Notten, P. H.~L., 2021.
\newblock ``A review of degradation mechanisms and recent achievements for ni‐rich cathode‐based li‐ion batteries''.
\newblock {\em Adv. Energy Mater., {\bf 11}}(48), Dec., p.~2103005.

\bibitem{Attia2022-jm}
Attia, P.~M., Bills, A., Planella, F.~B., Dechent, P., dos Reis, G., Dubarry, M., Gasper, P., Gilchrist, R., Greenbank, S., Howey, D., Liu, O., Khoo, E., Preger, Y., Soni, A., Sripad, S., Stefanopoulou, A.~G., and Sulzer, V., 2022.
\newblock ``{Review---``Knees''} in {Lithium-Ion} battery aging trajectories''.
\newblock {\em J. Electrochem. Soc., {\bf 169}}(6), June, p.~060517.

\bibitem{Brosa_Planella2022-aq}
Brosa~Planella, F., Ai, W., Boyce, A.~M., Ghosh, A., Korotkin, I., Sahu, S., Sulzer, V., Timms, R., Tranter, T.~G., Zyskin, M., Cooper, S.~J., Edge, J.~S., Foster, J.~M., Marinescu, M., Wu, B., and Richardson, G., 2022.
\newblock ``A continuum of physics-based lithium-ion battery models reviewed''.
\newblock {\em Prog. Energy Combust. Sci., {\bf 4}}(4), July, p.~042003.

\bibitem{Brosa_Planella2023-ae}
Brosa~Planella, F., and Widanage, W.~D., 2023.
\newblock ``A single particle model with electrolyte and side reactions for degradation of lithium-ion batteries''.
\newblock {\em Appl. Math. Model., {\bf 121}}, Sept., pp.~586--610.

\bibitem{Chen2020-og}
Chen, C.-H., Brosa~Planella, F., O'Regan, K., Gastol, D., Widanage, W.~D., and Kendrick, E., 2020.
\newblock ``Development of experimental techniques for parameterization of multi-scale lithium-ion battery models''.
\newblock {\em J. Electrochem. Soc., {\bf 167}}(8), p.~080534.

\bibitem{Wang2022-tl}
Wang, A.~A., O'Kane, S. E.~J., Brosa~Planella, F., Le~Houx, J., O'Regan, K., Zyskin, M., Edge, J., Monroe, C.~W., Cooper, S.~J., Howey, D.~A., Kendrick, E., and Foster, J.~M., 2022.
\newblock ``Review of parameterisation and a novel database ({LiionDB}) for continuum li-ion battery models''.
\newblock {\em Prog. Energy Combust. Sci., {\bf 4}}(3), May, p.~032004.

\bibitem{Rajamani2020-me}
Rajamani, R., Jeon, W., Movahedi, H., and Zemouche, A., 2020.
\newblock ``On the need for switched-gain observers for non-monotonic nonlinear systems''.
\newblock {\em Automatica, {\bf 114}}, Apr., p.~108814.

\bibitem{Boyd1994}
Boyd, S.~P., 1994.
\newblock {\em Linear Matrix Inequalities in System and Control Theory}.
\newblock Society for Industrial and Applied Mathematics.

\end{thebibliography}

%%%%%%%%%%%%%%%%%%%%%%%%%%%%%%%%%%%%%%%%%%%%%%%%%%%%%%%%%%%%%%%%%%%%%%
\appendix       %%% starting appendix
\section*{Appendix A: Nonlinear SOC Imbalance Bounds}
\label{sec:bounds-derivation}

From Eqs. (\ref{eqn:ib}-\ref{eqn:zdot}), the system can be represented as
\begin{align}
    \dot{z}_1(t) &= -\frac{I_1(t)}{Q_1} = \frac{+\Delta U(t) - I(t)R_2}{Q_1\rtot} \\
    \dot{z}_2(t) &= -\frac{I_2(t)}{Q_2} = \frac{-\Delta U(t) - I(t)R_1}{Q_2\rtot} 
\end{align}
where $\Delta U(t) \triangleq U(z_2(t)) - U(z_1(t))$. We can redefine the state as 
\begin{equation}
    x = \begin{bmatrix} x_1 \\ x_2 \end{bmatrix} 
      = \begin{bmatrix} \Delta z \\ Q_1z_1 + Q_2z_2 \end{bmatrix}.
\end{equation}
The dynamic system can then be presented as:
\begin{align}
    \label{eq:system}
    \dot{x}_1 &= \Delta \dot{z} = \mathcal{A}\Delta U(x_1,x_2) + \mathcal{B}I(t) \\
    \label{eq:dotx2}
    \dot{x}_2 &= -I(t)
\end{align}
where
\begin{align}
    \mathcal{A} &= -\frac{1}{\rtot}\left(\frac{1}{Q_1} + \frac{1}{Q_2}\right) \\
    \mathcal{B} &= \frac{1}{\rtot}\left(\frac{R_1}{Q_2} - \frac{R_2}{Q_1}\right)
\end{align}

\noindent Since OCV functions monotonically increase, the bounds on the slope of the OCV function can be defined as:

\begin{equation}
    \label{eq:ocv-bounds}
    0 < k_1 \leq \frac{\partial U(z)}{\partial z} \leq k_2
\end{equation}

\noindent Using the differential mean theorem from \cite{Rajamani2020-me}, we have 

\begin{equation}
    \label{eq:diffmean}
    k_1\Delta z \leq \Delta U \leq k_2 \Delta z.
\end{equation}

\noindent Note that $x_2$ in (\ref{eq:dotx2}) is neutrally stable. We therefore present the following definition for partial asymptotic stability.

\begin{definition}[Partial asymptotic stability \cite{Haddad2008}]
The general system 
\begin{align}
    \begin{split}
    \dot{x}_1 = f_1(x_1,x_2) \\
    \dot{x}_2 = f_2(x_1,x_2)
    \end{split}
\end{align}
is said to be asymptotically stable with respect to $x_1$ uniformly in initial value $x_2(0)$ if, for every value of $x_2(0)$, $x_1$ is stable and for every value of $||x_1(0)|| < \delta \Rightarrow \lim_{t\rightarrow\infty}x_1(t) = 0$.
\end{definition}

\noindent\textit{Proposition 1.} The unforced system ($I \equiv 0$) presented in (\ref{eq:system}) is asymptotically stable with respect to $x_1$ uniformly in $x_2(0)$.

\begin{proof}
Take the Lyapunov function candidate:
\begin{equation}
    V = P\Delta z^2
\end{equation}
where $P>0$. Taking the derivative, we have 
\begin{align}
    \label{eq:six}
    \begin{split}
    \dot{V} &= 2\mathcal{A}P\Delta z \Delta U \\
            &= \begin{bmatrix} \Delta z \\ \Delta U \end{bmatrix}^T
               \begin{bmatrix} 0 & \mathcal{A}P \\ \mathcal{A}P & 0 \end{bmatrix}
               \begin{bmatrix} \Delta z \\ \Delta U \end{bmatrix}
    \end{split}
\end{align}
The following sector condition can be easily derived from (\ref{eq:ocv-bounds}) \cite{Rajamani2020-me}:
\begin{align}
    \label{eq:seven}
    V_1 & = \begin{bmatrix} \Delta z \\ \Delta U \end{bmatrix}^T
          \begin{bmatrix} k_1k_2 & -\frac{k_1+k_2}{2} \\ -\frac{k_1 + k_2}{2} & 1 \end{bmatrix}
          \begin{bmatrix} \Delta z \\ \Delta U \end{bmatrix} \leq 0.
\end{align}
Using the S-procedure lemma \cite{Boyd1994} for (\ref{eq:six}) and (\ref{eq:seven}), if the following matrix inequality is satisfied, then $x_1$ is asymptotically stable \cite{Haddad2008}:
\begin{equation}
    \begin{bmatrix} -k_1k_1 & \mathcal{A}P + \frac{k_1 + k_2}{2} \\
                    \mathcal{A}P + \frac{k_1 + k_2}{2} & -1 \end{bmatrix}
                    \preceq 0.
\end{equation}
Since $k_1,k_2 > 0$ and $\mathcal{A} < 0$, this inequality is always satisfied.
\end{proof}

\begin{remark}
The same logic can be applied for multiple cells in parallel by taking $\Delta z$ for every two consecutive cells.
\end{remark}
Since, the system presented in (\ref{eq:system}) is input-$x_1$ stable, Theorem \ref{eq:theorem} holds, with the proof given below. 

\begin{proof}
    Here,  we follow the procedure for input-state stability similar to \cite{Khalil2002}. Since the unforced system is asymptotically stable with respect to $x_1$, we can write:
    \begin{align}
        \begin{split}
        \label{eq:w}
        \frac{d||\Delta z||}{dt} &= \frac{\Delta z \Delta \dot z}{||\Delta z||} \\
        &= \frac{\mathcal{A}\Delta U \Delta z + \mathcal{B} I \Delta z}{||\Delta z||}.
        \end{split}
    \end{align}

    \noindent Using the left hand side of the inequality from (\ref{eq:diffmean}), we can have the upper bound of (\ref{eq:w}) as:
    \begin{equation}
       \frac{d||\Delta z||}{dt} \leq k_1\mathcal{A}||\Delta z|| + |\mathcal{B}|||I(t)||.
    \end{equation}

    \noindent By using the comparison lemma \cite{Khalil2002}, we have:

    \begin{equation}
        \label{eq:inequality}
        ||\Delta z|| \leq |\Delta z(0)| e^{k_1\mathcal{A}t} + |\mathcal{B}|\int_0^te^{-(t-\tau)k_1\mathcal{A}}||I(t)||d\tau.
    \end{equation}
    which can be simplified as:
    \begin{equation}
        \label{eq:inequality}
        ||\Delta z|| \leq |\Delta z(0)| e^{k_1\mathcal{A}t} + |\mathcal{B}| \max(|I(t)|)\int_0^te^{-(t-\tau)k_1\mathcal{A}}d\tau.
    \end{equation}
    Taking the integral from 0 to $t$ results in (\ref{eq:linfbound}). It is apparent that condition (\ref{eq:condition}) guarantees that $||\Delta z|| \leq 1$.
\end{proof}

\renewcommand{\thefigure}{A\arabic{figure}}
\setcounter{figure}{0}

\begin{figure*}
\centering\includegraphics[width=\linewidth]{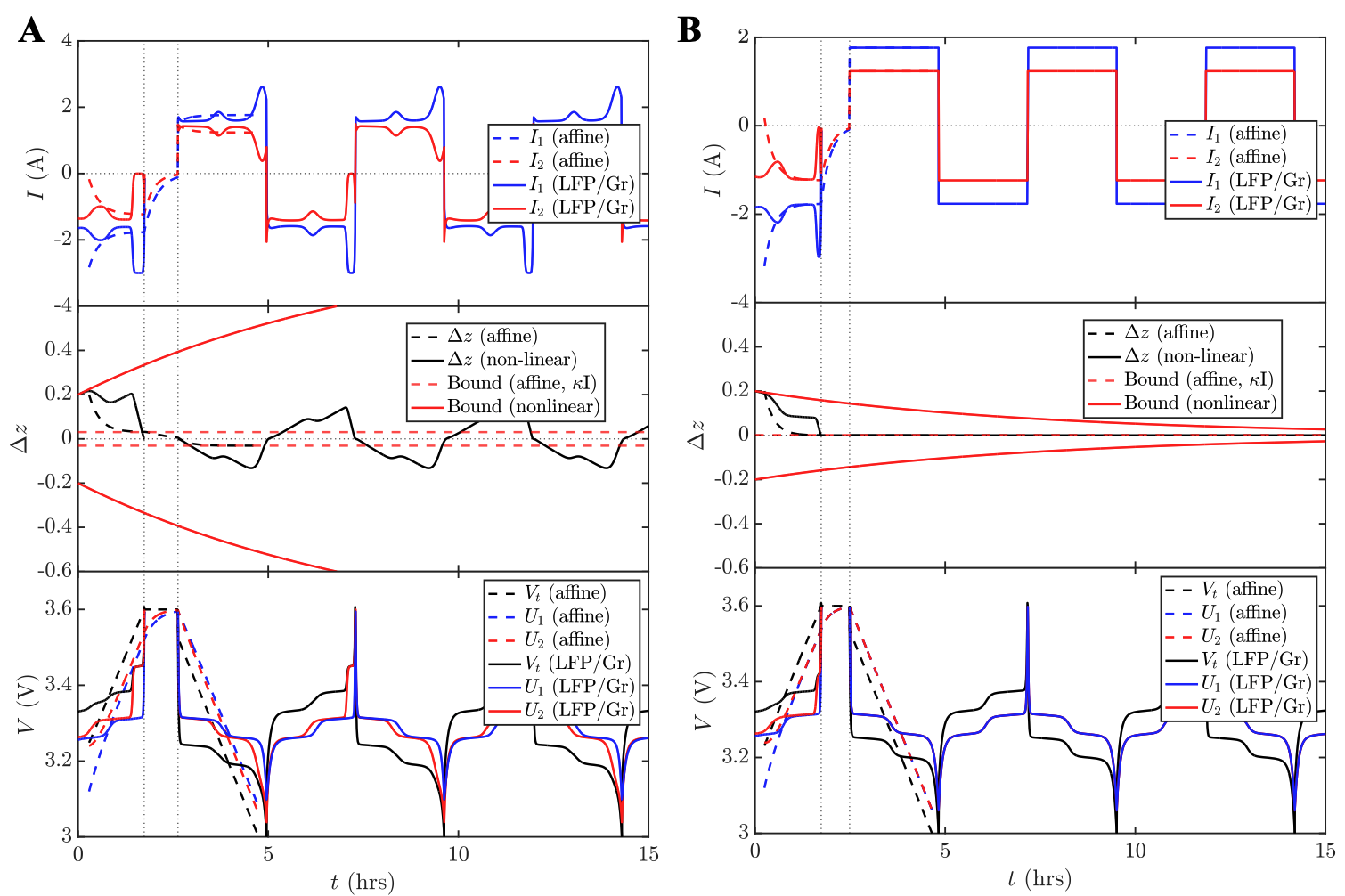}
\caption{\reva{OCV-R model system dynamics with nonlinear OCV functions} representing an LFP/Gr system. In both (A) and (B), $(Q_2,R_2)=(3$Ah$,50$m$\Omega$), $(z_{1,0},z_{2,0}) = (0.4, 0.2)$, $|I|=3$A during the constant current phases, and with a CV termination of $Q_2/5$ A. (A) uses $(Q_2/Q_1,R_2/R_1)=(0.7, 1.1)$. This pairing represents a typical scenario in which an aged cell (Cell 2), with lower capacity and higher resistance, is paired with a less aged cell (Cell 1). (B) uses $(Q_2/Q_1,R_2/R_1)=(0.7, 1.43)$. This pairing also represents a typical scenario with an aged cell paired with a less aged cell, but this pairing additionally satisfies the `$QR$-matching' condition, i.e. $Q_1R_1 = Q_2R_2$. Under this condition, the SOC imbalance dynamics become insensitive to the input current and exponentially decay to zero. The corresponding imbalance dynamics also become driven purely by resistance differences in the absence of SOC re-balancing currents. After the initial SOC imbalance fades, this system begins to behave identically to the affine OCV system, despite the presence of the nonlinear OCV function. See Section \ref{sec:qr-matching} for a complete discussion.} 
\label{fig:nonlinear-lfp}
\end{figure*}

\end{document}